\documentclass[11 pt]{article}
\usepackage[a4paper,top=3cm,bottom=3cm,left=2cm,right=2cm]{geometry} 

\usepackage{amsmath}
\usepackage{amssymb}
\usepackage{enumitem}
\usepackage{fontenc}
\usepackage{inputenc}
\usepackage{calc}
\usepackage{indentfirst}
\usepackage{fancyhdr}
\usepackage{graphicx}
\usepackage{epstopdf}
\usepackage{lastpage}
\usepackage{ifthen}
\usepackage{lineno}
\usepackage{float}
\usepackage{setspace}
\usepackage{mathpazo}
\usepackage{booktabs}
\usepackage{titlesec}
\usepackage{etoolbox}
\usepackage{tabto}
\usepackage{xcolor}
\usepackage{soul}
\usepackage{multirow}
\usepackage{microtype}
\usepackage{tikz}
\usepackage{totcount}
\usepackage{amsthm}
\usepackage{hyphenat}
\usepackage{hyperref}
\usepackage{footmisc}
\usepackage{url}
\usepackage{newfloat}
\usepackage{caption}

\usepackage{bbm}
\usepackage{amssymb}
\usepackage{mathtools}
\usepackage{subcaption}

\usepackage{quoting}

\begin{document}
\begin{center}
{\bf \Huge A continuous-time random walk extension of the Gillis model}
\end{center}

\vspace*{1 cm}
\noindent
{\bf Gaia Pozzoli $^{1,2}$, Mattia Radice $^{1,2}$, Manuele Onofri $^{1,2}$ and Roberto Artuso $^{1,2}$*}

\vspace*{0.6 cm}
{\noindent 
$^{1}$ \quad Dipartimento di Scienza e Alta Tecnologia and Center for Nonlinear and Complex Systems, Università degli Studi dell'Insubria, Via Valleggio 11, 22100 Como, Italy\\
$^{2}$ \quad I.N.F.N. Sezione di Milano, Via Celoria 16, 20133 Milano, Italy\\
\\
* \quad Correspondence: gpozzoli@uninsubria.it, m.radice1@uninsubria.it, m.onofri1@uninsubria.it, \\
roberto.artuso@uninsubria.it}

\vspace*{0.2 cm}
\begin{quote}
{{\bf Abstract:} We consider a continuous-time random walk which is the generalization, by means of the introduction of waiting periods on sites, of the one-dimensional nonhomogeneous random walk with a position-dependent drift known in the mathematical literature as \textit{Gillis random walk}. This modified stochastic process allows to significantly change local, non-local and transport properties in the presence of heavy-tailed waiting-time distributions lacking the first moment: we provide here exact results concerning hitting times, first-time events, survival probabilities, occupation times, the moments spectrum and the statistics of records. Specifically, normal diffusion gives way to subdiffusion and we are witnessing the breaking of ergodicity. Furthermore we also test our theoretical predictions with numerical simulations.}
\end{quote}

\begin{quote}
\end{quote}
\vspace*{0.4 cm}

\noindent
\today


\section{Introduction}
Since their first appearance, random walks have always been used as effective mathematical tools to describe a plenty of problems from a variety of fields, such as crystallography, biology, behavioural sciences, optical and metal physics, finance and economics. Although homogeneous random walks are not a mystery anymore, in many situations the topology of the environment causes correlations (induced by the medium inhomogeneities), which have powerful consequences on the transport properties of the process. The birth of whole classes of non-homogeneous random walks \cite{RWRE I, NHRW} is due to the need to study disordered media and non-Brownian motions, responsible for anomalous diffusive behaviour. This topic of research is prompted by phenomena observed in several systems such as turbolent flows, dynamical systems with intermittencies, glassy materials, Lorentz gases, predators hunting for food \cite{Anomalous Transport 1,Phys. Rep. 1,Nature 1,Adv Water Resour 1, Soil Sci Soc Am J 1,PLoS ONE 1, Phys. Rev. Lett. 2, Phys. Today 2, Phys. Chem. Chem. Phys. 1, Phys. Rev. Lett. 3, Phys. Chem. Chem. Phys. 2}. For the sake of clarity, we recall that anomalous processes are characterized by a mean square displacement of the walker's position with a sublinear or superlinear growth in time as opposed to normal Brownian diffusion, defined as an asymptotically linear time dependence of the variance.

In this context, the outstanding Gillis random walk \cite{Quart. J. Math. 1} plays a crucial role, since it is one of the few analytically solvable models of non-homogeneous random walks with a drift dependent on the position in the sample. Other few exceptions are random walks with a limited number of boundaries or defective sites \cite{ADV APPL PROBAB 1, J. Math. Phys. 2, J. Appl. Probab. 1,J. Stat. Phys. 4}. The Gillis model is a nearest-neighbour centrally-biased random walk on $\mathbbm{Z}$, lacking translational invariance for the transition probabilities, which provides an appropriate environment in order to investigate the critical behaviour in the proximity of a phase transition: while keeping fixed the dimensionality of the model, one can observe different regimes by simply changing the parameter value.

As is natural, in the first instance one typically focuses his attention on the dynamics of the random walk by considering a discretization of the time evolution: basically you wear the simplest clock, consisting of a counting measure of the number of steps. But in most physical situations, you deal with systems requiring a continuous-time description of the evolution (that clearly introduces a higher degree of complexity). In order to show this important difference, we can rely on the explanatory comparative analysis between Lévy flights and Lévy walks \cite{Rev. Mod. Phys. 1, On growth and form 1}. We are mentioning homogeneous random walks whose transition probabilities have an infinite variance: Lévy flights are indeed jump processes with steps picked from a long-tailed (or Lévy) distribution, whose tails are not exponentially bounded and so there is a good chance that you jump really far from the current site. This is what we mean when we say, from a mathematical point of view, that the distribution does not have a finite variance. However, Lévy flights have a drawback: clearly, if spatial trajectories are totally unaware of the related time trace, flights are rightful, otherwise they are not exactly physically acceptable because they appear to possess an infinite speed. More realistic models are instead Lévy walks, where the walker needs a certain time to perform the jump, which is no longer instantaneous. The time spent is usually proportional to the length of the step, so we assume a constant finite speed for the motion.

In our case, we take a step back: in Lévy walks you are already assuming the existence of spatiotemporal correlations, but in general the easiest way to get a continuous-time description starting from the discrete model is to decouple spatial and temporal components. This is precisely what E. Montroll and G.H. Weiss did in $1964$ \cite{J. Math. Phys. 1} by means of a random walk (the so called \textit{Continuous Time Random Walk}) whose jumps are still instantaneous but the dynamics is subordinated to a random physical clock. Basically, you have to introduce a new random variable, the waiting time on a site, in addition to the length of the jump \cite{First Steps in Random Walks}. Also this time there are relevant application aspects: ruin theory of insurance companies, dynamics of prices in financial markets, impurity in semiconductors, transport in porous media, transport in geological formations. An incomplete list of general references includes \cite{Physica A 1, Springer 1, Phys. Today 1, Rev. Geophys. 1, Water Resour. Res. 1, Water Resour. Res. 2}.

Inspired by the previous models, we consider the continuous-time generalization of the discrete-time Gillis random walk that we have already studied thoroughly in \cite{J. Stat. Mech. 1}. In particular, we will also look at first-time events: they account for key problems in the theory of stochastic processes, since you can determine \textit{when} system variables assume specific values (for example, see \cite{Cambridge 1}).

This paper is structured as follows. In Section~\ref{review} we briefly present the background in order to provide a complete overview of the known results, which are the basis of the work.  Then, in Section~\ref{results}, we will discuss the original results, by establishing a connection between the discrete-time random walk and the continuous-time formalism. In particular, two significant phenomena will arise: the ergodicity breaking and the extension of the anomalous diffusion regime. In Section~\ref{numerical results}, moreover, we will integrate the theoretical analysis with computational simulations, as further confirmation. Finally, in Section~\ref{discussion} we will summarize all the conclusions previously described in detail. 

\section{Review of previous work}\label{review}
First of all, intending to be self-consistent, we provide a brief recap of the discrete-time Gillis model and resume the key concepts necessary for its continuous-time version. In this way we will be sufficiently equipped to move on to list the major results. 
\subsection{Gillis random walk}
The Gillis model \cite{Quart. J. Math. 1} is a discrete-time random walk, on a one-dimensional lattice, whose transition probabilities $p_{i,j}$ of moving from site $i$ to site $j$ are non-null if and only if $|i-j|=1$, namely $i,j$ are nearest-neighbour lattice points. We assume that the positional dependence is ruled by the real parameter $\epsilon$, where $|\epsilon| <1$ and:

\begin{equation}
\mathcal{R}_j \coloneqq p_{j,j+1}= \frac 1 2 \left(1-\frac{\epsilon}{j}\right),\quad \mathcal{L}_j \coloneqq p_{j,j-1} = \frac 1 2 \left(1+\frac{\epsilon}{j}\right) \quad \mbox{for } j\in\mathbbm{Z}\setminus\{0\},\qquad \mathcal{R}_0\coloneqq\frac 1 2 \eqqcolon \mathcal{L}_0.
\end{equation}

Clearly, if you set out $\epsilon = 0$ you recover homogeneity since it boils down to the simple symmetric random walk. Otherwise, the position-dependent drift is responsible for an attractive bias towards the starting site, the origin, when $0 < \epsilon < 1$ or for a repulsive action if $-1<\epsilon<0$.

As we said in the general introduction, the Gillis random walk is one of the few analytically solvable models and, in particular, in his original paper the author writes down the exact expression of the generating function $P(z)$ of $\{p_n\}_{n\in\mathbbm{N}}$ in terms of the elementary hypergeometric function $_2 F_1(a,b;\:c;\:z)$ \cite{Handbook 1}, where $p_n\coloneqq p_n(0,0)$ denotes the probability that the walker returns to the origin, not necessarily for the first time, after $n$ steps.  Actually, this solution has been later generalized for a generic starting site \cite{RWRE I}. Given the probability $p_n(j_0,j=0)$ that the particle starts from any site $j_0$ and passes through the origin after $n$ steps, we can write its generating function:

\begin{equation}\label{pr-hughes}
P(j_0,0;\: z)=\sum_{n=0}^{\infty}p_{2n+|j_0|} (j_0,0)z^{2n+|j_0|}=\frac{z^{|j_0|}}{|j_0|!}\frac{\Gamma(1+\epsilon+|j_0|)}{2^{|j_0|}\Gamma(1+\epsilon)}\frac{_{2}F_1\left( \frac{\epsilon + 1+|j_0|}{2}, \frac{ \epsilon+|j_0|}{2}+1;\: |j_0|+1;\: z^2 \right)}{_{2}F_1\left( \frac 1 2 \epsilon, \frac 1 2 \epsilon + \frac 1 2;\: 1;\;  z^2 \right)}.
\end{equation}

This is one of the essential tools for the following analysis, along with those in our previous paper \cite{J. Stat. Mech. 1}, and for $j_0=0$ it is clearly consistent with the original result by Gillis concerning the generating function $P(z)\coloneqq P(0,0;\:z)$.

Another relevant statement for future considerations is that the motion is positive-recurrent (recurrent with a finite mean return time) and ergodic (thus admitting a stationary distribution) when $\frac 1 2 < \epsilon < 1$, null-recurrent (recurrent with an infinite mean return time that increases with the number of steps) if $-\frac 1 2 \leq \epsilon \leq\frac 1 2$ and transient for $- 1 <\epsilon < -\frac 1 2$. To be more precise \cite{Physica A 2}, the mean time taken between two consecutive returns to the starting site up to the $n$-th step is: 

\begin{equation}
\langle\tau_{ret}(n)\rangle\sim \begin{cases}
n^{3/2+\epsilon} \qquad &\mbox{if}\quad -1<\epsilon<-\frac1 2,\\
\frac{n}{\ln^2(n)} \qquad &\mbox{if}\quad \epsilon=-\frac 1 2,\\
n^{1/2+\epsilon}\qquad &\mbox{if}\quad -\frac 1 2<\epsilon<+\frac 1 2,\\
\ln(n) \qquad& \mbox{if}\quad \epsilon=+\frac 1 2,\\
\frac{2\epsilon}{2\epsilon-1}\qquad&\mbox{if} \quad+\frac 1 2<\epsilon<+1,
\end{cases}
\end{equation}
and this is a direct consequence of Eq.~\eqref{pr-hughes}. In fact, starting from there one can also obtain the generating functions of the first-hitting and the first-return times to the origin. First of all, let us define the probability $f_n(j_0,j)$ that the moving particle starts from $j_0$ and hits $j$ for the first time after $n$ steps. Then we know that $\{f_n(j_0,j)\}_{n\in\mathbbm{N}}$ are connected to $\{p_n(j_0,j)\}_{n\in\mathbbm{N}}$ in the following way:

\begin{equation}\label{rec-first-ret}
p_n(j_0,j)=\delta_{n,0}\delta_{j,j_0}+\sum_{k=1}^n f_k(j_0,j)p_{n-k}(j,j),
\end{equation} 
or, equivalently, in terms of the corresponding generating functions: 

\begin{equation}\label{F-P}
F(j_0,j;\:z)=\frac{P(j_0,j;\:z)-\delta_{j,j_0}}{P(j,j;\:z)}.
\end{equation}

Notice that in the presence of translational invariance $p_{n-k}(j,j)=p_{n-k}$ and $P(j,j;\:z)=P(z)$ (see Eq.~$(2.8)$ in \cite{First Steps in Random Walks}). In our context, anyway, Eq.~\eqref{F-P} becomes particularly easy since we choose $j=0$. Hence we can finally conclude that when $j_0= 0$:

\begin{equation}\label{eq-f-r-0}
F(z)\coloneqq F(0,0;\:z) =\sum_{n=1}^{\infty} f_{2n}z^{2n}=1-\frac{1}{P(z)}= 1-\frac{_2 F_1\left( \frac 1 2 \epsilon, \frac 1 2 \epsilon + \frac 1 2; \: 1; \: z^2 \right)}{_2F_1\left( \frac 1 2 \epsilon+\frac 1 2, \frac 1 2 \epsilon+1; \: 1; \: z^2\right)},
\end{equation}
where $f_n\coloneqq f_n(0,0)$ are the first-return probabilities, whereas for $j_0\neq 0$ (first-passage probabilities):

\begin{multline}\label{eq-f-r}
F(j_0,0;\: z)=\sum_{n=0}^{\infty}f_{2n+|j_0|} (j_0,0)z^{2n+|j_0|}= \frac{P(j_0,0;\: z)}{P(z)}\\
=\frac{z^{|j_0|}}{|j_0|!}\frac{\Gamma(1+\epsilon+|j_0|)}{2^{|j_0|}\Gamma(1+\epsilon)}\frac{_2F_1\left( \frac{\epsilon+1+|j_0|}{2},\frac{\epsilon+|j_0|}{2}+1;\:|j_0|+1;\:z^2\right)}{_2F_1\left( \frac 1 2 \epsilon +1, \frac{\epsilon+1}{2};\:1;\:z^2\right)}.
\end{multline}

The mean time spent between two consecutive visits at the origin up to the $n$-th step is easily derived from $\langle \tau_{ret}(n)\rangle = \lim_{z\to 1^-}\frac{F'(z)}{F(z)}$ \cite{Physica A 2}.

Now, instead, moving on to transport properties, we can quickly resume the moments spectrum and the statistics of records (for more details and references see \cite{J. Stat. Mech. 1}). Firstly, denoting the moment of order $q$ with $\langle|j_n|^q\rangle\coloneqq\sum_{j\in\mathbbm{Z}}p_n(0,j)|j|^q$, we know that the asymptotical dependence on the number of steps $n$ is $\langle|j_n|^q\rangle\sim n^{\nu_q}$, where:

\begin{equation}\label{gillis-moments}
\nu_q=\nu_q(\epsilon)=\begin{cases}
\frac{q}{2} \qquad & \mbox{if} \quad \epsilon <\frac 1 2,\\
0 \qquad & \mbox{if} \quad \epsilon>\frac 1 2 \quad \mbox{and}\quad q<2\epsilon -1,\\
\frac{1+q}{2}-\epsilon \qquad & \mbox{if}\quad \epsilon>\frac 1 2 \quad \mbox{and}\quad q>2\epsilon -1.
\end{cases}
\end{equation}

Translated into words, this leads to recognize the presence of a phase transition: non-ergodic processes are characterized by normal diffusion, since the second moment shows an asymptotically linear growth in time, whereas the ergodic ones reveal strong anomalous (sub-)diffusion \cite{Physica D 1}.

Secondly, let us first recall the following definition: given a finite set of random variables, the record value is the largest/smallest value assumed by that sequence. In the Gillis model, the events to account for are the positions $\{j_k\}_{k\in\mathbbm{N}}$ on the one-dimensional lattice during the motion and the record after $n$ steps $R_n$, with $R_0=0$ due to the intial condition $j_0=0$, is the non-negative integer exceeding all the previously occupied sites. In addition, the presence of a nearest-neighbour structure implies that the number of records after $n$ steps $N_n$ is connected to the value of the maximum $M_n\coloneqq\max_{1\leq k\leq n}j_k$ by means of the trivial relationship $N_n=M_n+1$, where:

\begin{equation}
\langle M_n\rangle \sim \begin{cases}
n^{1/2} \qquad & \mbox{if}\quad -\frac1 2 <\epsilon \leq + \frac 1 2,\\
n^{1/(1+2\epsilon)}\qquad & \mbox{if} \quad +\frac 1 2\leq \epsilon <+1.
\end{cases}
\end{equation}

Here, again, the model enters two different phases, according to the value of the characteristic parameter $\epsilon$. In particular, in the interval $\epsilon\in \left(-\frac 1 2, \frac 1 2 \right)$ the mean number of records has the same growth of the first moment.

\subsection{CTRW}
Our aim here is to formalize the transformation of the number of steps into the physical real time. We shall follow \cite{First Steps in Random Walks}. As a preliminary remark, we point out that, moving from discrete to continuous formalism, we have to abandon the generating function (for the time domain, not for the lattice) in favour of a more appropriate mathematical tool, the Laplace transform.

The basic assumption is that we have a random walker who performs instantaneous jumps on a line, but now he is forced to wait on the target site for a certain interval of time, whose duration $t$ is always picked from a common probability distribution $\psi(t)$, before going any further. So, for instance, $t_1$ will be the waiting time on the origin before jumping for the first time and, moreover, we would emphasize that the waiting times of subsequent steps are \textit{independent and identically distributed} (according to $\psi(t)$) random variables.

These are the essential instruments for introducing the quantities of interest. Firstly, we can define the \textit{Probability Density Function} $\psi_n(t)$ of the occurrence of the $n$-th step at time $t=t_1+\dots+t_n$. As a consequence, through independence, the following recurrence relation holds:

\begin{equation}
\psi_n(t)= \int_0^t \psi_{n-1}(t')\psi(t-t')dt'\qquad \implies \qquad \hat{\psi}_n(s)=\hat{\psi}_{n-1}(s)\hat{\psi}(s)=\dots=\hat{\psi}^n(s),
\end{equation}
where the convolution becomes a product and, from now on, the use of the following convention is implied: variables indicated in brackets in the functions uniquely define the space you are working in (real for $t$, Laplace for $s$). Secondly, we can introduce the PDF $\chi_n(t)$ of taking exactly $n$ steps up to the time $t$ (namely this time the $n$-th step may occur at time $t'<t$ and then the walker rests on the site):

\begin{equation}
\chi_n(t)=\int_0^t\psi_n(t')\underbrace{\left[1-\int_0^{t-t'}\psi(\tau)d\tau\right]}_{\substack{\mbox{survival probability}\\ \mbox{on a site } \chi_0(t-t')}}dt'  \qquad \implies \qquad \hat{\chi}_n(s)=\hat{\psi}^n(s)\frac{1-\hat{\psi}(s)}{s}.
\end{equation}

Next section will shed some light on the role of these useful quantities.
 
\section{Results}\label{results}
For the sake of clarity, we will simply state the significant elements in this section. For all of the detailed computation please refer to appendices further below.

\subsection{Probability of being at the origin}
The most natural step to undertake as first thing is obviously to determine the probability of finding the walker at the origin at time $t$, for comparison with Gillis original results. This task can be carried out in two different ways, both instructive.

\subsubsection{Gillis way}\label{sec-gillis-way}
As a first attempt, one could be led to translate Gillis method into continuous-time formalism. And in fact this is a viable solution.
The starting point is Eq.~$(2.1)$ in \cite{Quart. J. Math. 1} that reads:

\begin{equation}\label{discrete-gillis}
p_{n+1}(j)=p_n(j-1) \mathcal{R}_{j-1}+p_n(j+1)\mathcal{L}_{j+1},
\end{equation}
where $p_n(j)\coloneqq p_n(0,j)$ denotes the probability of being at site $j$ after $n$ steps when the initial position of the walker is the origin.

In order to accomplish the transformation, we need to establish some more notation. In particular we notice that, after introducing the physical real time, the position at time $t$ is still the position after $n$ steps, provided that exactly $n$ jumps have been counted up to time $t$. Hence:

\begin{itemize}[leftmargin=*,labelsep=5.8mm]
\item $p(j,t)=\sum_{n=0}^{\infty}p_n(j)\chi_n(t)=\int_0^tp^a(j,t')\chi_0(t-t')dt'$ is the probability of \textit{being} (arriving) at $j$ at (within) time $t$;
\item $p^a(j,t)=\sum_{n=0}^{\infty}p_n(j)\psi_n(t)$ is the probability of \textit{ arriving} at $j$ at time $t$.
\end{itemize}

 By performing the Laplace transform on time and the generating function on sites, we get: $\hat{P}(x,s)\coloneqq\int_0^{\infty}dt e^{-st}\sum_{j=0}^{\infty}p(j,t)x^j=\hat{P}^a(x,s)\cdot \hat{\chi}_0(s)= \hat{P}^a(x,s)\frac{1-\hat{\psi}(s)}{s}$. Now, the continuous-time equivalent of Eq.~\eqref{discrete-gillis} is obtained by multiplying both sides by $\psi_{n+1}(t)$ and summing over $n$:

\begin{equation}
\sum_{n=0}^{\infty}p_{n+1}(j)\psi_{n+1}(t)=
\begin{cases}
\sum_{n=0}^{\infty}p_n(j)\psi_n(t)-p_0(j)\psi_0(t)=p^a(j,t)-\delta_{j,0}\delta(t),\\
\int_0^t p^a(j-1,t')\psi(t-t')dt'\mathcal{R}_{j-1}+\int_0^tp^a(j+1,t')\psi(t-t')dt' \mathcal{L}_{j+1}.
\end{cases}
\end{equation}

Essentially, we find ourselves in the exact same situation, we just need to shift focus back to a new key element, the arrival event. Retracing the steps of the original paper (see Appendix~\ref{app-gillis-way}), we can (trivially) conclude that:

\begin{equation}\label{gillis-res}
\hat{p}^a(j=0;\: s)=\frac{\int_0^{2\pi}(1-\hat{\psi}(s)\cos\theta)^{-1-\epsilon}d\theta}{\int_0^{2\pi}(1-\hat{\psi}(s)\cos\theta)^{-\epsilon}d\theta}=\frac{_{2}F_1\left( \frac 1 2 \epsilon + 1, \frac 1 2 \epsilon + \frac 1 2;\: 1;\: \hat{\psi}^2(s) \right)}{_{2}F_1\left( \frac 1 2 \epsilon, \frac 1 2 \epsilon + \frac 1 2;\: 1;\: \hat{\psi}^2 (s)\right)},
\end{equation}
namely:

\begin{equation}\label{prob-being}
\hat{p}(s)\coloneqq\hat{p}(j=0;\: s)=\frac{1-\hat{\psi}(s)}{s}\hat{p}^a(j=0;\: s)=\frac{1-\hat{\psi}(s)}{s}P[z=\hat{\psi}(s)],
\end{equation}
where we remind you that $P[\hat{\psi}(s)]$ is the generating function (evaluated at $\hat{\psi}(s)$) of the probability of being at the origin in the discrete-time model.

This result is not surprising: given that the temporal component is independent of the spatial scale, the time trace is ruled by a random clock that replaces the role of the counting measure (the simple internal clock given by the number of steps). The generating function of the probability of \emph{arriving} at the origin is the same of the one associated with the discrete model (where there is no distinction between arriving and being, because the walker can not stand still on a site), but subordinated to the new physical time. This observation let us generalize immediately the result to the case of a generic starting point $j_0$, obtaining:

\begin{equation}
\hat{p}(j_0,0;\: s)=\frac{1-\hat{\psi}(s)}{s}\hat{p}^a(j_0,0;\:s)=\frac{1-\hat{\psi}(s)}{s}P[j_0,0;\: z=\hat{\psi}(s)],
\end{equation}
which, thanks to Eq.~\eqref{pr-hughes}, becomes:

\begin{equation}\label{pr-hitting-laplace}
\hat{p}(j_0,0;\: s)=\frac{1-\hat{\psi}(s)}{s}\left(\frac{\hat{\psi}(s)}{2}\right)^{|j_0|}\frac{\Gamma(1+\epsilon+|j_0|)}{|j_0|!\Gamma(1+\epsilon)}\frac{_{2}F_1\left( \frac{\epsilon + 1+|j_0|}{2}, \frac{ \epsilon+|j_0|}{2}+1;\: |j_0|+1;\: \hat{\psi}^2 (s)\right)}{_{2}F_1\left( \frac 1 2 \epsilon, \frac 1 2 \epsilon + \frac 1 2;\: 1;\;  \hat{\psi}^2(s) \right)}.
\end{equation}

\subsubsection{Recurrence relation: first-return time to the origin}
However, we can arrive at Eq.~\eqref{prob-being} also in a different way.  If we now perform, as before, a continuous-time transformation of Eq.~\eqref{rec-first-ret} with $j_0=0=j$, we get:

\begin{equation}
p(t)=\chi_0(t)+\sum_{n=1}^{\infty}\sum_{k=1}^n f_kp_{n-k}\chi_n(t),
\end{equation}
and considering the Laplace domain:

\begin{eqnarray}
\hat{p}(s) 
& = & \frac{1-\hat{\psi}(s)}{s}\left(1+\sum_{n=1}^{\infty}\sum_{k=1}^nf_kp_{n-k}\hat{\psi}^n(s)\right)\\
& = & \frac{1-\hat{\psi}(s)}{s}\left(1+F[z=\hat{\psi}(s)]P[z=\hat{\psi}(s)]\right).
\end{eqnarray}

As a last step we can plug in Eq.~\eqref{F-P}, so we finally go back to Eq.~\eqref{prob-being}. In addition, we have immediately also the Laplace transform for the first-return time. Indeed, since the first-return is an arrival event and thus coincides with the occurrence of a step, there is no way to land earlier and wait for the remaining time. Hence, from a mathematical point of view, we can write \cite{First Steps in Random Walks}:

\begin{equation}\label{prob-first}
f(t):=f(j=0,t)=\sum_{n=0}^{\infty}f_n\psi_n(t) \qquad \implies \qquad \hat{f}(s)=F[z=\hat{\psi}(s)]=1-\frac{1}{P[z=\hat{\psi}(s)]},
\end{equation}
thanks to Eq.~\eqref{F-P}. Lastly, by comparing Eq.~\eqref{prob-being} with Eq.~\eqref{prob-first}, we get the relationship in the Laplace domain:

\begin{equation}\label{eq-f-r-0-CT}
\hat{f}(s)=1-\frac{1-\hat{\psi}(s)}{s\hat{p}(s)}.
\end{equation}

Once again we can generalize the previous formula for a generic starting site $j_0\neq 0$: 

\begin{equation}\label{eq-f-r-CT}
\hat{f}(j_0,0;\: s)=F[j_0,0;\: z=\hat{\psi}(s)]=\frac{P[j_0,0;\:z=\hat{\psi}(s)]}{P[z=\hat{\psi}(s)]}=\frac{\hat{p}(j_0,0;\:s)}{\hat{p}(s)}.
\end{equation}

Now, turning back to our specific case, we know that the generating functions of interest can be written in the form (see \cite{J. Stat. Mech. 1}):

\begin{eqnarray}
P(z) & = & \frac{1}{(1-z)^{\rho}}H\left(\frac{1}{1-z}\right),\\
F(z) & = & 1-(1-z)^{\rho}L\left(\frac{1}{1-z}\right), \qquad \mbox{for}\quad \epsilon\geq -\frac 1 2,
\end{eqnarray}
where $L(x)=1/H(x)$ are slowly-varying functions at infinity, namely for instance $L:\mathbbm{R}_+ \to \mathbbm{R}_+$ is such that $\forall c>0$ $\exists \lim_{x\to \infty}\frac{L(cx)}{L(x)}=1$, and:

\begin{equation}
\rho=\begin{cases} 0\quad&\mbox{if} \: -1<\epsilon \leq -\frac 1 2\\ 
\frac 1 2 +\epsilon \quad& \mbox{if} \: -\frac 1 2 <\epsilon <+\frac 1 2 \\
 1 \quad & \mbox{if}\:+\frac 1 2 \leq \epsilon <+ 1. 
\end{cases}
\end{equation}

As a consequence, the corresponding Laplace transforms are automatically given by:

\begin{eqnarray}
\hat{p}(s) & = & \frac{[1-\hat{\psi}(s)]^{1-\rho}}{s}H\left(\frac{1}{1-\hat{\psi}(s)}\right),\\
\hat{f}(s) & = &  1 - [1-\hat{\psi}(s)]^{\rho}L\left(\frac{1}{1-\hat{\psi}(s)}\right).
\end{eqnarray}

At this point, it is apparent that we are forced to split our analysis according to the features of the waiting-time distribution: clearly the asymptotic behaviour of the quantities above mentioned is established by the expansion of the Laplace transform of the waiting-time distribution $\hat{\psi}(s)$ for small $s$.

\subsubsection{Finite-mean waiting-time distributions}\label{sec-fin-mean}
As the first choice, one can think of $\{t_i\}_{i\in\mathbbm{N}}$ as i.i.d. positive random variables with finite mean $\tau$ (but non necessarily a finite variance too: for example, the waiting times may be taken belonging to the domain of attraction - because they must be spectrally positive \cite{arXiv 1} - of $\alpha$-stable laws with $\alpha\in(1,2)$). In these circumstances, the leading term in the expansion is: $\hat{\psi}(s)=1-\tau s+o(s)$ for $s\to 0$. Therefore, in the limit $s\to 0$ we get:

\begin{align}
\hat{p}(s)  \sim  \frac{\tau^{1-\rho}}{s^{\rho}}H\left(\frac{1}{\tau s}\right)\qquad &\implies\qquad p(t)\sim \frac{1}{\Gamma(\rho)}\frac{\tau^{1-\rho}}{t^{1-\rho}}H\left(\frac{t}{\tau}\right),\quad&\mbox{with}\quad 0<\rho\leq 1, \\
\hat{f}(s)  \sim   1 - \tau^{\rho}s^{\rho}L\left(\frac{1}{\tau s}\right)\qquad& \implies \qquad f(t)\sim \frac{\rho}{\Gamma(1-\rho)}\frac{\tau^{\rho}}{t^{1+\rho}}L\left(\frac{t}{\tau}\right), \quad &\mbox{with}\quad 0<\rho< 1,
\end{align}
equivalently written in the limit $t\to\infty$ by directly applying Tauberian theorems \cite{Feller II,First Steps in Random Walks}. In any case, however, the exponent of the power-law decay is the same of the discrete-time model \cite{J. Stat. Mech. 1}:

\begin{equation}
p_{2n}\sim \begin{cases}
n^{-\frac 1 2+\epsilon}\qquad &\mbox{if}\quad -1<\epsilon<+\frac 1 2,\\
\frac{4}{\ln(n)} \qquad &\mbox{if}\quad \epsilon = +\frac 1 2,\\
2-\frac{1}{\epsilon}\qquad &\mbox{if}\quad +\frac 1 2<\epsilon<+1,
\end{cases} \qquad \qquad f_{2n}\sim\begin{cases}
n^{-\frac 1 2+\epsilon}\qquad &\mbox{if}\quad -1<\epsilon<-\frac 1 2,\\
\frac{1}{n\ln^2(n)} \qquad &\mbox{if}\quad \epsilon = -\frac 1 2,\\
n^{-\frac 3 2-\epsilon}\qquad &\mbox{if}\quad -\frac 1 2<\epsilon<+1.
\end{cases}
\end{equation}

This is not an astonishing result because obviously $t\sim \tau n$, where $n$ is the number of steps. It is merely a change of scale. So from now on we will disregard this possibility.

\subsubsection{Infinite-mean waiting-time distributions}\label{sec-inf-mean}
Implications are a little bit different if we choose power-law distributions lacking the first moment, because the dynamics becomes highly irregular this time. If we assume a heavy-tailed waiting-time distribution of the form: $\psi(t)\sim\frac{B}{t^{1+\alpha}}$ with $0<\alpha<1$, then the corresponding Laplace expansion is:  $\hat{\psi}(s)=1-b s^{\alpha}+o(s^{\alpha})$, where $b\coloneqq B\frac{\Gamma(1-\alpha)}{\alpha}$ . Again by substitution, we derive:

\begin{align}\label{taub-asymp-1}
\hat{p}(s)  \sim  \frac{b^{1-\rho}}{s^{1-\alpha(1-\rho)}}H\left(\frac{1}{bs^{\alpha}}\right)\qquad &\implies\qquad p(t)\propto \frac{1}{t^{\alpha(1-\rho)}}, \\ \label{taub-asymp-2}
\hat{f}(s)  \sim   1 - b^{\rho}s^{\alpha\rho}L\left(\frac{1}{bs^{\alpha}}\right)\qquad& \implies \qquad f(t)\propto \frac{1}{t^{1+\alpha\rho}},  \quad& \mbox{with}\quad 0< \rho \leq 1.
\end{align}

We invite you to read Appendix~\ref{app-CTRW-example} for a check with a well-known example. Moreover, as advanced in the previous section, asymptotic expansion and Tauberian theorems give us an immediate, even if incomplete, insight of what happens: exact and exhaustive results (involving a generic starting site) are postponed in Appendices~\ref{app-exact-ret} and \ref{app-exact-first-ret}, in order not to burden the reading. 

Anyhow, we provide here the summarising full spectrum of return, first-return, hitting and first-hitting time PDFs of the origin, which is consistent with the asymptotic behaviour previously predicted in Eq.~\eqref{taub-asymp-1} and for $\epsilon > -\frac 1 2$ in Eq.~\eqref{taub-asymp-2}. Firstly, keeping in mind Eq.~\eqref{pr-hitting-laplace}, we have:

\begin{equation}\label{asympt-ret-0}
p(j_0,0,t)\sim \begin{cases}
\frac{b}{-2\epsilon-1}\frac{\Gamma(1+\epsilon+|j_0|)\Gamma(1-\epsilon)}{\Gamma(1+\epsilon)\Gamma(|j_0|-\epsilon)}\frac{1}{\Gamma(1-\alpha)}\frac{1}{t^{\alpha}}\qquad &\mbox{if}\quad -1<\epsilon<-\frac 1 2,\\
\frac 1 4\ln\left( \frac{t^{\alpha}}{2b}\right)\frac{b}{\Gamma(1-\alpha)}\frac{1}{t^{\alpha}} \qquad &\mbox{if}\quad \epsilon = -\frac 1 2 ,\\
\left(\frac{b}{2}\right)^{\frac 1 2-\epsilon} \frac{\Gamma(1-\epsilon)\Gamma\left(\frac 1 2+\epsilon\right)}{\Gamma\left(\frac 1 2-\epsilon\right)\Gamma(1+\epsilon)}\frac{1}{\Gamma\left(1-\frac{\alpha}{2}+\alpha\epsilon\right)}\frac{1}{t^{\alpha\left(\frac 1 2-\epsilon\right)}}\qquad & \mbox{if}\quad -\frac 1 2 <\epsilon<+\frac 1 2,\\
2 \left[\ln\left(\frac{t^{\alpha}}{2b}\right)\right]^{-1}\qquad & \mbox{if}\quad\epsilon=+\frac 1 2,\\
1-\frac{1}{2\epsilon}\qquad & \mbox{if}\quad+\frac 1 2 <\epsilon<+1.
\end{cases}
\end{equation}

With the choice $j_0=0$, you get immediately the PDF of returns. In particular, let us point out that in the recurrent cases the coefficient does not depend on $j_0$.

A little discrepancy, instead, arises if you compare first-passage and first-return events (see Eq.~\eqref{eq-f-r-0} and Eq.~\eqref{eq-f-r}). The first-return time PDF is asymptotically given by:

\begin{equation}\label{asympt-first-ret-0}
f(t)\sim \begin{cases}
2^{\epsilon-\frac 1 2}\left( \frac{2\epsilon+1}{\epsilon}\right)^2\frac{\Gamma\left(\frac 3 2 +\epsilon\right)\Gamma(1-\epsilon)}{\Gamma(\epsilon+1)\Gamma\left(\frac 1 2-\epsilon\right)}\frac{\alpha}{\Gamma\left(1+\alpha\left(\frac 1 2 +\epsilon\right)\right)}\frac{b^{-\frac 1 2-\epsilon}}{t^{1-\alpha\left(\frac 1 2+\epsilon\right)}}\qquad &\mbox{if}\quad -1<\epsilon<-\frac 1 2,\\
\frac{4\alpha}{\ln^2\left(\frac{t^{\alpha}}{2b} \right)}\frac 1 t\qquad &\mbox{if}\quad \epsilon = -\frac 1 2 ,\\
2^{\frac 1 2-\epsilon} \frac{\Gamma\left(\frac 1 2-\epsilon\right)\Gamma(1+\epsilon)}{\Gamma(1-\epsilon)\Gamma\left(\frac 1 2+\epsilon\right)}\frac{\alpha\left(\frac 1 2 +\epsilon\right)}{\Gamma\left(1-\frac{\alpha}{2}-\alpha\epsilon\right)}\frac{b^{\frac 1 2+\epsilon}}{t^{1+\alpha\left(\frac 1 2 +\epsilon \right)}}\qquad & \mbox{if}\quad -\frac 1 2 <\epsilon<+\frac 1 2,\\
\frac{b}{2} \ln\left(\frac{t^{\alpha}}{2b}\right)\frac{\alpha}{\Gamma(1-\alpha)}\frac{1}{t^{1+\alpha}} \qquad & \mbox{if}\quad\epsilon=+\frac 1 2,\\
\frac{2\epsilon}{2\epsilon-1}\frac{\alpha}{\Gamma(1-\alpha)}\frac{b}{t^{1+\alpha}}\qquad & \mbox{if}\quad+\frac 1 2 <\epsilon<+1,
\end{cases}
\end{equation}
whereas the first-hitting time PDF can be connected to the previous one by means of the following relationship:

\begin{equation}
f(j_0,0,t)\sim C_{\epsilon}(j_0)f(t),
\end{equation}
with:

\begin{equation}\label{eq-coeff}
C_{\epsilon}(j_0)=\begin{cases}
\frac{1}{2\epsilon+1}\left[\epsilon+\frac{\Gamma(1+\epsilon+|j_0|)\Gamma(1-\epsilon)}{\Gamma(1+\epsilon)\Gamma(|j_0|-\epsilon)}\right]\qquad &\mbox{if}\quad -1<\epsilon<-\frac 1 2,\\
\frac1 4 \left[\Psi\left(\frac 1 4 \right) + \Psi \left(\frac 3 4 \right)-\Psi\left(\frac 1 4 +\frac{|j_0|}{2}\right)-\Psi\left(\frac 3 4 +\frac{|j_0|}{2}\right)\right]\qquad&\mbox{if}\quad \epsilon=-\frac 1 2,\\
\frac{1}{2\epsilon+1}\left[\epsilon+\frac{\Gamma(1+\epsilon+|j_0|)\Gamma(1-\epsilon)}{\Gamma(1+\epsilon)\Gamma(|j_0|-\epsilon)}\right]\qquad &\mbox{if}\quad -\frac 1 2<\epsilon<+\frac 1 2,\\
j_0^2 \qquad &\mbox{if}\quad \epsilon=+\frac 1 2,\\
\frac{j_0^2}{2\epsilon} \qquad &\mbox{if}\quad +\frac 1 2<\epsilon<+1,
\end{cases}
\end{equation}
and $\Psi(z)$ denoting the digamma function \cite{Handbook 1}. As a consequence, we notice that, by setting $j_0=0$, the asymptotic expansion of $f(j_0,0,t)$ vanishes in all regimes: the direct relation between $p(j_0,0,t)$ and $p(t)$ does not hold anymore for first-time events. Nevertheless, although the coefficients differ, the asymptotic decays of $f(j_0,0,t)$ and $f(t)$ are the same.

\subsection{Survival probability on the positive semi-axis}
Return and first-return probabilities allow us to determine also the asymptotic behaviour of other related quantities. In the first place, we can introduce the survival probability in a given subset: it is defined as the probability $q_n$ of never escaping from the selected collection of neighbouring sites. For instance, by considering $\mathbbm{N}$, it can be written as:

\begin{equation}
q_n\coloneqq\mathbb{P}(j_1\geq 0,j_2\geq 0,\dots,j_n\geq 0|j_0=0),\qquad q_0\coloneqq1.
\end{equation}

This quantity has been deepenly studied for a wide range of homogeneous stochastic processess. In particular, with regards to random walks with i.i.d. steps, the historical Sparre-Andersen theorem \cite{Math. Scand. 1} is a significant result connecting a non-local property, since the survival probability depends on the history of the motion, to the local (in time) probability of standing non-positive at the last step:

\begin{equation}
Q(z)=\sum_{n=0}^{\infty}q_nz^n=\exp\left\{\:\sum_{n=1}^{\infty}\frac{z^n}{n}\mathbb{P}(j_n\leq 0)\:\right\}, \qquad q_n\sim n^{-1/2}.
\end{equation}

It is an outstanding expression of universality, both in discrete and continuous-time versions, if you consider jump distributions that are continuous and symmetric about the origin, although this feature is partially lost (the coefficient of proportionality in the scaling law is no longer universal) when you move on a lattice instead of the real line \cite{J. Phys. A Math. Theor. 1, Feller I}. However, whereas temporal components have already been included in the analysis \cite{Phys. Rev. E 2}, not much has been said about spatial correlations, to the authors' knowledge.

With our previous results \cite{Phys. Rev. E 1} in mind, we will now consider changes arising from the subordination to a physical clock. We need to introduce the persistence probability $u_n$ of never coming back to the origin up to the $n$-th step, namely $u_n:=1-\sum_{k=0}^{n}f_k=\mathbb{P}(j_1\neq 0,j_2\neq 0,\dots,j_n\neq 0|j_0=0)$, in order to write down the following recurrence relation:

\begin{equation}
2q_n=\delta_{n,0}+u_n+\sum_{k=1}^n f_k q_{n-k}\qquad \implies \qquad 2q(t)=\chi_0(t)+u(t)+\sum_{n=1}^{\infty}\sum_{k=1}^n f_k q_{n-k}\chi_n(t).
\end{equation}

In the Laplace domain it becomes:

\begin{equation}
2\hat{q}(s)=\hat{u}(s)+\frac{1-\hat{\psi}(s)}{s}\left(1+F[\hat{\psi}(s)]Q[\hat{\psi}(s)]\right), \qquad \mbox{where}\quad \hat{u}(s)=\frac{1-F[\hat{\psi}(s)]}{s},
\end{equation}
and in conclusion:

\begin{equation}
\hat{q}(s)\approx \frac{1}{s^{1-\alpha\rho}},\quad s\to 0\qquad \implies \qquad q(t)\approx \frac{1}{t^{\alpha\rho}}, \quad t\to \infty,
\end{equation}
to be compared with the discrete-time results \cite{Phys. Rev. E 1}:

\begin{equation}
Q(z)=\frac{1+U(z)}{1+(1-z)U(z)}, \qquad U(z)=\frac{1-F(z)}{1-z}=\frac{1}{(1-z)^{1-\rho}}L\left(\frac{1}{1-z}\right), 
\end{equation}
with $Q(z)\sim U(z)$ as $z\to 1$ and $-\frac 1 2 < \epsilon < \frac 1 2$. It is apparent that there are similarities in the null-recurrent cases $-\frac 1 2 < \epsilon<\frac 1 2 $, since $q_n\sim n^{-\rho}$, and when $\epsilon \leq - \frac 1 2$, whose decay is even slower (for $\epsilon=-\frac 1 2$ it is a decreasing slowly-varying function). The main relevant difference, instead, is the disappearance of the positive-recurrent regime (where Tauberian theorems fail).

Nevertheless, if the underlying discrete-time random walk is ergodic, a discrepancy remains also in the continuous-time translation. In general, we have to notice that:

\begin{equation}\label{surv-halved}
\hat{q}(s)=\hat{u}(s)-\frac{s\hat{u}^2(s)}{2}+\frac{1-\hat{\psi}(s)}{2s}\sim\begin{cases}
\left[1-\frac 1 2 L\left(\frac{1}{bs^{\alpha}}\right)\right]\hat{u}(s) \qquad & -1<\epsilon<-\frac 1 2,\\
\hat{u}(s) \qquad & -\frac 1 2\leq\epsilon\leq+\frac 1 2,\\
 \left[1+\frac {1}{2L}\right]\hat{u}(s) \qquad & + \frac 1 2 <\epsilon<+1,
\end{cases}
\end{equation}
where the slowly-varying function $L$ tends to a constant. The coefficient of proportionality between $q(t)$ and $u(t)$ should not be underestimated: it means that the occupation time spent at the origin behaves in a different way, as we will see later on. Indeed, for the sake of illustration, a halved coefficient $q(t)\sim\frac 1 2 u(t)$ would mean that visits at the origin are negligible (which is recovered in the limiting case $\epsilon=-1$). In particular, $1+\frac {1}{2L}>1$ says that they have more weight in the presence of an underlying ergodic context.

As a final check, if we suppose, on the contrary, a finite time scale ($\hat{\psi}(s)\sim 1-\tau s$) then we have again $q(t)\sim u(t)$ (independently of $\tau$, clearly) and $q(t)\sim t^{-\rho}$ when $-\frac 1 2<\epsilon <\frac 1 2$.

\subsection{Occupation times}\label{sec-occ-time}
This section will be devoted to the statistics of the fraction of time spent by the walker at a given site or in a given subset. As in the previous one, the probability distribution of the quantity of interest stems from the features of the asymptotic decay of return and first-return PDFs. We shall describe (or simply mention) any other necessary tool from time to time, as always.

\subsubsection{Occupation time of the origin}
In the discrete-time formalism, as we have already discussed in Section~\ref{sec-gillis-way}, the particle can not stand still on a site and so considering the occupation time of a single site is equivalent to talk about the number of visits to the same. Thanks to the Darling-Kac theorem \cite{Trans. Am. Math. Soc. 1}, a remarkable mathematical result for Markov processes, we know \cite{Phys. Rev. E 1,J. Stat. Mech. 1} that the number of visits to the starting point (properly rescaled by the average taken over several realizations) has a Mittag-Leffler distribution of index $\rho$ as limiting distribution. We would emphasize that spatial inhomogeneities cause non-Markovianity for the original process, but now we are focusing on returns to the origin that are renewal events. Thus you have a sequence of i.i.d. first-return times and loss of memory is ensured in each case. 

Obviously, this result is still true for waiting-time distributions with finite mean: even if  the physical clock is running when the particle rests on a site and the internal clock stops, the microscopic time scale gives us the constant of direct proportionality necessary to move from the number of steps to the correct time measure, which has the same distribution, as a consequence.

In the non-trivial continuous-time translation, instead, what we need to apply is Lamperti theorem \cite{Trans. Am. Math. Soc. 2}. It is a statement involving two-state stochastic processes (being or not at the origin in our case). More precisely we deal with its continuous-time generalization, which has been discussed in many works, such as \cite{J. Stat. Phys. 2, J. Stat. Phys. 3, J. Phys. Condens. Matter 1, Phys. Rev. E 3, Phys. Rev. Lett. 1}. Here we provide the final formula, that is the starting point of our analysis: for a detailed proof refer to \cite{J. Phys. Condens. Matter 1}, for instance. Essentially, we will conclude that, even if the Mittag-Leffler statistics is mapped to a Lamperti distribution, the index $\rho$ of the discrete-time formalism is always replaced by the product $\alpha\rho$ characterizing the asymptotic expansion of the first-return PDF. In particular, in order to preserve the ergodic property of the discrete-time version ($\rho=1$) we have to consider a waiting-time distribution with finite mean ($\alpha=1$): in this way, the Lamperti distribution collapses to a Dirac delta function on the mean value of the occupation time.

Before formalizing the theorem, let us define some notation. We consider a stochastic process described by a set of transitions between two states (that we call $in$ and $out$) and we consider arrivals at the origin and departures as events. Time periods between events are i.i.d. random variables, with PDFs $\psi_{in}(t)\equiv \psi(t)$ and $\psi_{out}(t)\equiv f(1,0,t)$ respectively, that are the alternating distributions of the renewal process. In fact, the time spent on state $in$ is precisely the waiting time on a site, whereas the time spent outside the origin coincides with the first-return time to the origin starting from $j_0=1$ because, thanks to the nearest-neighbour structure, when you leave the origin you land on $\pm 1$ and $f(j_0,j,t)=f(|j_0|,j,t)$ by symmetry, as witnessed by Eq.~\eqref{eq-coeff}. Moreover we can notice that $\psi_{in}$ and $\psi_{out}$ are connected by means of the first-return PDF, in fact:

\begin{equation}
f(t)=\int_0^t \psi(t')\psi_{out}(t-t')dt' \qquad \implies \qquad \hat{f}(s)=\hat{\psi}(s)\hat{\psi}_{out}(s).
\end{equation}

We assume that at $t=0$ the particle occupies the origin (namely it is in state $in$) and we denote the total times spent by the walker in the two states up to time $t$ by $T_{in}$ and $T_{out}$, associated with the PDFs $f_t^{in}(T_{in})$ and $f_t^{out}(T_{out})$. Continuous-time Lamperti theorem tells us that the double Laplace transforms of these quantities are:

\begin{eqnarray}\label{CTlamp}
\hat{f}_s^{in}(u) &=& \left[\hat{\psi}_{in}(s+u)\frac{1-\hat{\psi}_{out}(s)}{s}+\frac{1-\hat{\psi}_{in}(s+u)}{s+u}\right]\frac{1}{1-\hat{\psi}_{in}(s+u)\hat{\psi}_{out}(s)},\\
\hat{f}_s^{out}(u) &=& \left[\hat{\psi}_{in}(s)\frac{1-\hat{\psi}_{out}(s+u)}{s+u}+\frac{1-\hat{\psi}_{in}(s)}{s}\right]\frac{1}{1-\hat{\psi}_{in}(s)\hat{\psi}_{out}(s+u)}.
\end{eqnarray}

For the moment we focus on the non-ergodic regime $\epsilon \leq \frac 1 2$. First of all, let us choose a finite-mean waiting-time distribution, which constitutes a useful check. Clearly $\hat{\psi}_{in}(s)=\hat{\psi}(s)\sim 1-\tau s$ and we know that $\hat{f}(s)\sim 1-\tau^{\rho}s^{\rho}L\left(\frac{1}{\tau s}\right)$: having different asymptotic time decays, $\hat{\psi}_{out}$ is ruled by the slower one, namely $\hat{\psi}_{out}(s)\sim\hat{f}(s)$ (and indeed $C_{\epsilon}(1)=1$ as you can see from Eq.~\eqref{eq-coeff}). By substituting in Eq.~\eqref{CTlamp}, we immediately get:

\begin{equation}
\hat{f}_s^{in}(u)\sim \frac{\tau+\tau^{\rho}s^{\rho-1}L\left(\frac{1}{\tau s}\right)-\tau^{\rho+1}s^{\rho-1}(s+u)L\left(\frac{1}{\tau s}\right)}{\tau(s+u)+\tau^{\rho}s^{\rho}L\left(\frac{1}{\tau s}\right)-\tau^{\rho+1}s^{\rho}(s+u)L\left(\frac{1}{\tau s}\right)},
\end{equation}
and by expanding in powers of $u$, one can compute the moments of order $k$ of $T_{in}(t)$ in the time domain:

\begin{multline}
\langle T_{in}^k (s)\rangle=(-1)^k\frac{\partial^k}{\partial u^k}\hat{f}_s^{in}(u)\Big|_{u=0}\sim \frac{k!}{L^k\left(\frac{1}{\tau s}\right)}\frac{\tau^{k(1-\rho)}}{s^{1+k\rho}},\quad s \to 0\\
\implies \qquad \langle T_{in}^k(t)\rangle \sim \frac{k!}{\Gamma(1+k\rho)}\frac{\tau^{k(1-\rho)}}{L^k\left(\frac{t}{\tau }\right)}t^{k\rho}, \quad t \to \infty.
\end{multline}

This suggests to us that if we consider the rescaled random variable:

\begin{equation}\label{eq-ML}
\zeta(t)\coloneqq\frac{L\left(\frac{t}{\tau}\right)}{\tau^{1-\rho}}\frac{T_{in}(t)}{t^{\rho}} \qquad \implies \qquad \lim_{t\to\infty}\mathbb{E}[ \zeta^k(t)] = \frac{\Gamma(1+k)}{\Gamma(1+k\rho)},
\end{equation}
then we asymptotically recover the moments of the Mittag-Leffler function of index $\rho$, as we said previously. We would point out that $\zeta$ is not directly the fraction of time spent at the origin and this observation is consistent with the fact that, in addition to the presence of an infinite recurrence time, $f(t)$ decays more slowly with respect to $\psi(t)$: without a properly scaling, $T_{in}(t)$ is negligible with respect to $T_{out}(t)$ and, from a mathematical point of view, it follows a Dirac delta with mass at the origin, namely all moments converge to $0$ .

If now, instead, we take waiting-time distributions with infinite mean, we can not find out any scaling function in such a way that the rescaled occupation time admits a limiting distribution. In fact, recalling that $\hat{\psi}(s)\sim 1-bs^{\alpha}$ and $\hat{f}(s)\sim 1-b^{\rho}s^{\alpha\rho}L\left(\frac{1}{bs^{\alpha}}\right)$, we similarly obtain: 

\begin{eqnarray}
\hat{f}_s^{in}(u)&\sim& \frac{b(s+u)^{\alpha-1}+b^{\rho}s^{\alpha\rho-1}L\left(\frac{1}{b s^{\alpha}}\right)-b^{\rho+1}s^{\alpha\rho-1}(s+u)^{\alpha}L\left(\frac{1}{b s^{\alpha}}\right)}{b(s+u)^{\alpha}+b^{\rho}s^{\alpha\rho}L\left(\frac{1}{b s^{\alpha}}\right)-b^{\rho+1}s^{\alpha\rho}(s+u)^{\alpha}L\left(\frac{1}{b s^{\alpha}}\right)},\\
\langle T_{in}^k(t)\rangle & \sim & \frac{(-1)^{k+1}k\Gamma(\alpha)b^{1-\rho}t^{k+\alpha(\rho-1)}}{L\left(\frac{t^{\alpha}}{b}\right)\Gamma(\alpha-k+1)\Gamma(1+k+\alpha(\rho-1))}\quad \implies \quad \lim_{t\to \infty}\mathbb{E}\left[\left( \frac{T_{in}(t)}{t}\right) ^k\right]=0.
\end{eqnarray}

Let us move on to the discrete-time ergodic regime: $\epsilon > \frac 1 2$ and $\rho =1$. This time $\hat{\psi}(s)$ and $\hat{f}(s)$ are of the same order, since they possess the same asymptotic exponent and the slowly-varying function decays to a constant $L$. As a consequence, they both determine the behaviour of:

\begin{equation}
\hat{\psi}_{out}(s)=\hat{f}(1,0;\:s)\sim 1-(L-1)bs^{\alpha},
\end{equation}
according to $C_{\epsilon}(1)=\frac{1}{2\epsilon}$.

By exploiting again Eq.~\eqref{CTlamp}, in the limit $s\to 0$ we have:

\begin{equation}
\hat{f}^{in}_s(u)\sim\frac 1 s \frac{\left(1+\frac u s \right)^{\alpha-1}+L-1}{\left(1+\frac u s \right)^{\alpha}+L-1},
\end{equation}
which may be inverted (see \cite{Princeton 1} as in the original paper \cite{Trans. Am. Math. Soc. 2}) and leads to the Lamperti probability density function for the fraction of time $\frac{T_{in}(t)}{t}$ spent at the origin (ergodicity breaking):

\begin{equation}\label{eq-lamp-occ-time}
G'_{\eta,\alpha}(t)=\frac{a\sin(\pi\alpha)}{\pi}\frac{t^{\alpha-1}(1-t)^{\alpha-1}}{a^2t^{2\alpha}+2at^{\alpha}(1-t)^{\alpha}\cos(\pi\alpha)+(1-t)^{2\alpha}},
\end{equation}
where $a=L-1$ is the asymmetry parameter and $\eta\coloneqq\lim_{t \to \infty}\mathbb{E}\left(\frac{T_{in}(t)}{t}\right)=\frac 1 L$. In addition, we notice that:

\begin{equation}
\langle\tau_{ret}\rangle\coloneqq\sum_{n=1}^{\infty}nf_n=\lim_{z\to1^-}F'(z)=L=\frac{2\epsilon}{2\epsilon-1},
\end{equation}
and so the expected value of the fraction of continuous-time spent at the origin coincides with the inverse mean recurrence time of the discrete-time random walk. But, thanks to ergodicity, we have also a stationary distribution $\pi_0$ at the origin for $\epsilon > \frac 1 2$ \cite{J. Stat. Mech. 1} that, by means of Birkhoff ergodic theorem, satisfies: 

\begin{equation}
\frac{1}{\langle \tau_{ret}\rangle} = \lim_{n\to\infty} \frac{\sum_{k=1}^n \delta_{j_k,0}}{n}=\langle \delta_{j,0}\rangle_t\stackrel{B}{=}\langle \delta_{j,0}\rangle_{ens}=\pi_0,
\end{equation}
and in conclusion:

\begin{equation}
a=\frac{1-\pi_0}{\pi_0}=\frac{\pi_{out}}{\pi_{in}},
\end{equation}
where $\pi_{out},\pi_{in}$ are the stationary measures of the subsets associated with the two states, according to the known results in the literature \cite{J. Stat. Phys. 3, J. Phys. Condens. Matter 1, Phys. Rev. E 3, Phys. Rev. Lett. 1}.

As a last comment, we turn back again to the finite-mean case. As expected, when $\alpha=1$ we get:

\begin{equation}
\hat{f}_s^{in}(u)\sim \frac{1}{s+\eta u},\qquad \lim_{t\to\infty}\mathbb{E}\left[\left(\frac{T_{in}(t)}{t}\right)^k\right]=\eta^k \qquad \implies \qquad f_t^{in}(T_{in})=\delta(T_{in}-\eta t),
\end{equation}
namely a Dirac delta centered at the expected value $\eta$.

\subsubsection{Occupation time of the positive semi-axis}
In the non-ergodic (for the discrete-time random walk) regime, since $\frac{T_{out}(t)}{t}\to 1$ given that the fraction of time spent at the origin is negligible (as we discussed above, after Eq.~\eqref{eq-ML}), we have a system with a state space split into two subsets, $\mathbbm{Z}_+$ and $\mathbbm{Z}_-$, that can communicate only passing through the recurrent event, the origin, that is also the initial condition. Thanks to symmetry, $\psi_{\mathbbm{Z}_+}(t)=\psi_{\mathbbm{Z}_-}(t)\sim \psi_{out}(t)$ and the limiting distribution of the fraction of time spent in each subset is the symmetric Lamperti PDF of index $\alpha\rho$, $G'_{\frac 1 2, \alpha\rho}$, which for finite-mean waiting times consistently boils down to $G'_{\frac 1 2, \rho}$ (by directly applying the original Lamperti statement \cite{Trans. Am. Math. Soc. 2}).

In the ergodic regime, instead, when you split the state $\mathbbm{Z}\setminus \{0\}$ in two symmetric subsets, you must in any case look at a three-state process: although the mean recurrence time is still infinite, the fraction of time spent at the origin has its weight  without any rescaling, see Eq.~\eqref{eq-lamp-occ-time}. But by symmetry you know also that $\frac{T_{\mathbbm{Z}_+}(t)}{t}=\frac 1 2 \frac{T_{out}(t)}{t}$: as a consequence, you can easily conclude that the Lamperti distribution is $G'_{\eta_+,\alpha}$ with $\eta_+=\frac{\eta_{out}}{2}=\frac{L-1}{2L}$. In fact, you can retrace previous steps for the asymptotic expansion of $\hat{f}_s^{out}(u)$ in Eq.~\eqref{CTlamp} or equivalently observe that $\mathbb{E}\left[\frac{T_{out}(t)}{t}\right]=1-\mathbb{E}\left[\frac{T_{in}(t)}{t}\right]=\frac{L-1}{L}=\frac{1}{2\epsilon}$ and the exponent $\alpha$ remains unchanged when you move from $\psi_{in}(t)$ to $\psi_{out}(t)$. And here too, the asymmetry parameter could be written as:

\begin{equation}
a=\frac{1-\pi_{\mathbbm{Z}_+}}{\pi_{\mathbbm{Z}_+}}=\frac{\pi_{\mathbbm{Z}_-\cup \{0\}}}{\pi_{\mathbbm{Z}_+}}.
\end{equation}

By way of conclusion, as in the previous section, if you set $\alpha=1$ then you obviously recover the ergodicity in the continuous-time model, since you get a Dirac delta with mass at $\eta_+$. Apparently this time there is a little difference with respect to the discrete-time random walk (see \cite{J. Stat. Mech. 1}): the degenerate distribution is no longer centered at $\frac 1 2$ (obtained immediately from Lamperti theorem \cite{Trans. Am. Math. Soc. 2}), as expected by symmetry. But this value was due to the convention \cite{Trans. Am. Math. Soc. 2} of counting the visits at the origin $\left (\frac{T_{in}(t)}{t}\not\to 0\right)$ according to the direction of motion. So, if we consider, in addition to the occupation time of the positive axis, half the time spent at the origin (in the long-time limit), then we correctly get a mass at $\eta_++\frac{\eta}{2}=\frac 1 2$. This comment allows us to highlight another aspect of the ergodicity breaking: when $\alpha<1$, on the contrary, the choice of the convention to be adopted is completely irrelevant to the final result, since the mean return time to the origin is infinite, supporting the asymmetry of the distribution.

\subsection{Moments spectrum}
Having assumed the presence of a waiting-time distribution of the form specified in Section~\ref{sec-inf-mean} and knowing the asymptotic behaviour of the moments with respect to the number of steps, Eq.~\eqref{gillis-moments}, all we have to do is find out the number of steps performed (on average) up to time $t$ in order to determine the physical time dependence of the moments.  Clearly we can write \cite{First Steps in Random Walks} $\langle n(t)\rangle = \sum_{n=0}^{\infty}n\chi_n(t)$ that in the Laplace domain reads:

\begin{equation}
\langle \hat{n}(s)\rangle = \frac{1-\hat{\psi}(s)}{s}\sum_{n=0}^{\infty} n \hat{\psi}^n(s)=\frac{1-\hat{\psi}(s)}{s}\cdot \hat{\psi}(s)\frac{d}{d\hat{\psi}(s)}\sum_{n=0}^{\infty}\hat{\psi}^n(s)=\frac{\hat{\psi}(s)}{s[1-\hat{\psi}(s)]}\sim \frac{1}{bs^{\alpha+1}}.
\end{equation}

Now, by applying Tauberian theorems once more and coming back to the time domain, we get:

\begin{equation}
\langle n(t)\rangle \sim \frac{1}{\Gamma(1+\alpha)}\frac{t^{\alpha}}{b}.
\end{equation}

As a consequence, we can easily conclude that:

\begin{equation}
\langle |j|^q(t)\rangle=\sum_{j\in\mathbbm{Z}}p(j,t)|j|^q\sim t^{\alpha\nu(q)}=\begin{cases}
t^{\frac{q}{2}\alpha} \qquad & \mbox{if} \quad \epsilon <\frac 1 2,\\
t^0 \qquad & \mbox{if} \quad \epsilon>\frac 1 2 \quad \mbox{and}\quad q<2\epsilon -1,\\
t^{\frac{1+q-2\epsilon}{2}\alpha} \qquad & \mbox{if}\quad \epsilon>\frac 1 2 \quad \mbox{and}\quad q>2\epsilon -1,
\end{cases}
\end{equation}
hence, in particular, a subdiffusive regime also arises for non-ergodic processes. The derivation of this spectrum for the discrete-time model is rather technical: it is a consequence of the specific form of the continuum limit. So we will not dwell on a brief recap this time, for the detailed analysis refer to \cite{J. Stat. Mech. 1,J. Stat. Phys. 1}.

\subsection{Statistics of records}
The statistics of records is another aspects relying on the mean number of steps counted in a given time period. Essentially, we have to retrace the relevant steps shown in \cite{J. Stat. Mech. 1} for the discrete-time random walk in the light of additional knowledge.

First we must outline an excursion as each subsequence between consecutive returns to the origin: as we shall see, properties of single excursions carry information about the expected value of the maximum of the entire motion. We have handled with a stochastic process defined on the half-line, for instance on the non-negative integers $\mathbbm{N}$: in the case of symmetric random walks, we do not need to deal with its extension to the whole line, the origin can always be assumed to be a totally reflecting barrier. Indeed changes take over if and only if positive and negative excursions are characterized by different tail bounds for their durations \cite{Stoch. Process. Their Appl. 1}. A fundamental assumption to fulfill, instead, is the presence of the regenerative structure, whereas Markovianity is not required. Moreover, in order to make sure that there is recurrence, we focus on the range $\epsilon>-\frac 1 2$.

For the sake of completeness, here we provide heuristic guidelines: they should simply be intended as a motivation, for rigourous proofs we entrust you to previous references. Let $E_n$ denote the number of excursions, equivalently the number of returns to the origin, occurred up to the $n$-th step and $M$ the maximum position occupied during a single excursion. In \cite{J. Stat. Mech. 1} we have shown that the stochastic process obtained from the Gillis random walk $\{j_k\}_{k\in\mathbbm{N}}$ by means of the transformation $j_n^{1+2\epsilon}$ is a symmetric random walk with no longer drift.  As a consequence, thanks to classic results in random walk theory (see \cite{Cambridge 1}), we know that the probability of reaching the site $m$ before coming back to the origin, that is also the probability of having $M$ beyond $m$, is given by:

\begin{equation}
\mathbb{P}(\mbox{hitting } m \mbox{ before going back to } 0) \sim \mathbb{P}(M\geq m)\sim \frac{1}{m^{1+2\epsilon}},
\end{equation}
and then:

\begin{equation}
\mathbb{P}(M_n<m)=[\mathbb{P}(M<m)]^{E_n}=\left(1-Cm^{-1-2\epsilon}\right)^{E_n},
\end{equation}
since, because of the renewal property, excursions are independent of one another. Now, by means of the common limits for exponential functions:

\begin{equation}
\lim_{n\to \infty}\left(1-\frac{x^{-1-2\epsilon}}{E_n}\right)^{E_n}=e^{-x^{-1-2\epsilon}},
\end{equation}
since recurrence ensures $E_n\to \infty$ as $n\to \infty$, we deduce that the correct scaling law for the maximum is $M_n\sim E_n^{\frac{1}{1+2\epsilon}}$. At this point, we have just to find out the relationship between the number of excursions $E_n$ and the number of steps $n$. But this is almost immediate since we know that the properly rescaled random variable $\frac{E_n}{n^{\rho}}$ follows a Mittag-Leffler distribution of parameter $\rho$ \cite{Phys. Rev. E 1,Trans. Am. Math. Soc. 1}, whose first moment is by definition $\frac{1}{\Gamma(1+\rho)}$ and as a consequence $\langle E_n\rangle \sim n^{\rho}$. In conclusion, we get that the expected value of the maximum reached by the particle up to time $t$ is:

\begin{equation}
\langle M_n\rangle \sim \begin{cases}
n^{\frac 1 2}\qquad & \mbox{if}\quad -\frac1 2 <\epsilon \leq + \frac 1 2,\\
n^{\frac{1}{1+2\epsilon}}\qquad & \mbox{if} \quad +\frac 1 2\leq \epsilon <+1,
\end{cases}\qquad \implies \qquad \langle M(t)\rangle \sim \begin{cases}
t^{\frac{\alpha}{2}} \qquad & \mbox{if}\quad -\frac1 2 <\epsilon \leq + \frac 1 2,\\
t^{\frac{\alpha}{1+2\epsilon}}\qquad & \mbox{if} \quad +\frac 1 2\leq \epsilon <+1.
\end{cases}
\end{equation}

An interesting comment concerns a related quantity, the duration of a single excursion $T$, namely the first-return time to the origin. A mathematical rigorous theorem \cite{Stoch. Process. Their Appl. 1} comes to our aid once again. On the event that $\{j_n\}_{n\in\mathbbm{N}}$ reaches $m$ during an excursion, semimartingale estimates can be used to show that approximately the walker spends an amount of time of order $m^2$ before returning to the origin:

\begin{equation}
\mathbb{P}(T>m^2)\sim \mathbb{P}(M>m) \qquad \implies \qquad \mathbb{P}(T>n)\sim n^{-\frac 1 2-\epsilon},
\end{equation}
that is clearly consistent with our result in Section~\ref{sec-fin-mean}: $f_{2n}\sim n^{-\frac 3 2 -\epsilon}$. Moreover, the expected value of the maximum duration of an excursion up to the $n$-th step is:

\begin{equation}
\langle T_n^{max}\rangle\sim \langle E_n\rangle^{\frac{2}{1+2\epsilon}}\sim\begin{cases}
n\qquad& \mbox{if}\quad -\frac 1 2<\epsilon \leq +\frac 1 2,\\
n^{\frac{2}{1+2\epsilon}}\qquad & \mbox{if}\quad +\frac 1 2\leq\epsilon < +1,
\end{cases}
\end{equation}
according to the fact that for $\epsilon \leq \frac 1 2$ the process is null-recurrent, whereas in the ergodic regime we have a finite mean return time and the growth of $\langle T_n^{max}\rangle$ is slower. On the contrary, as we have seen in Section~\ref{sec-occ-time}, in the presence of a non-trivial continuous-time random walk ergodicity is lost and in fact:

\begin{equation}
f(t)\sim t^{-1-\alpha\rho},\qquad \langle T^{max}(t)\rangle\sim t.
\end{equation} 



\section{Numerical results}\label{numerical results}
Here our intent is to substantiate theoretical arguments by means of numerical checks. Moreover, we also take the opportunity to show  how detailed analytical considerations are fundamental in this kind of context: some aspects are intrinsically difficult to be directly investigated from a numerical point of view.

Before going any further, as a general comment, from now on we will consider Pareto distributions as heavy-tailed waiting-time distributions for our simulations:

\begin{equation}
\psi(t)=\frac{\alpha t_0^{\alpha}}{t^{\alpha+1}},\qquad t>t_0,
\end{equation}
where $\alpha$ is a positive parameter, the so-called tail index, and $t_0$, the scale parameter, is the lower bound for $t$. In this way, the variance of the random variable for $\alpha \in (1,2]$ is infinite, with a finite mean, whereas it does not exist for $\alpha \leq 1$, when the expected value becomes infinite. We will focus on the latter case.

\subsection{Return and first-return events}
\begin{figure}[H]
	\centering
	\begin{subfigure}{.45\textwidth}
		\centering
		\includegraphics[width=\linewidth]{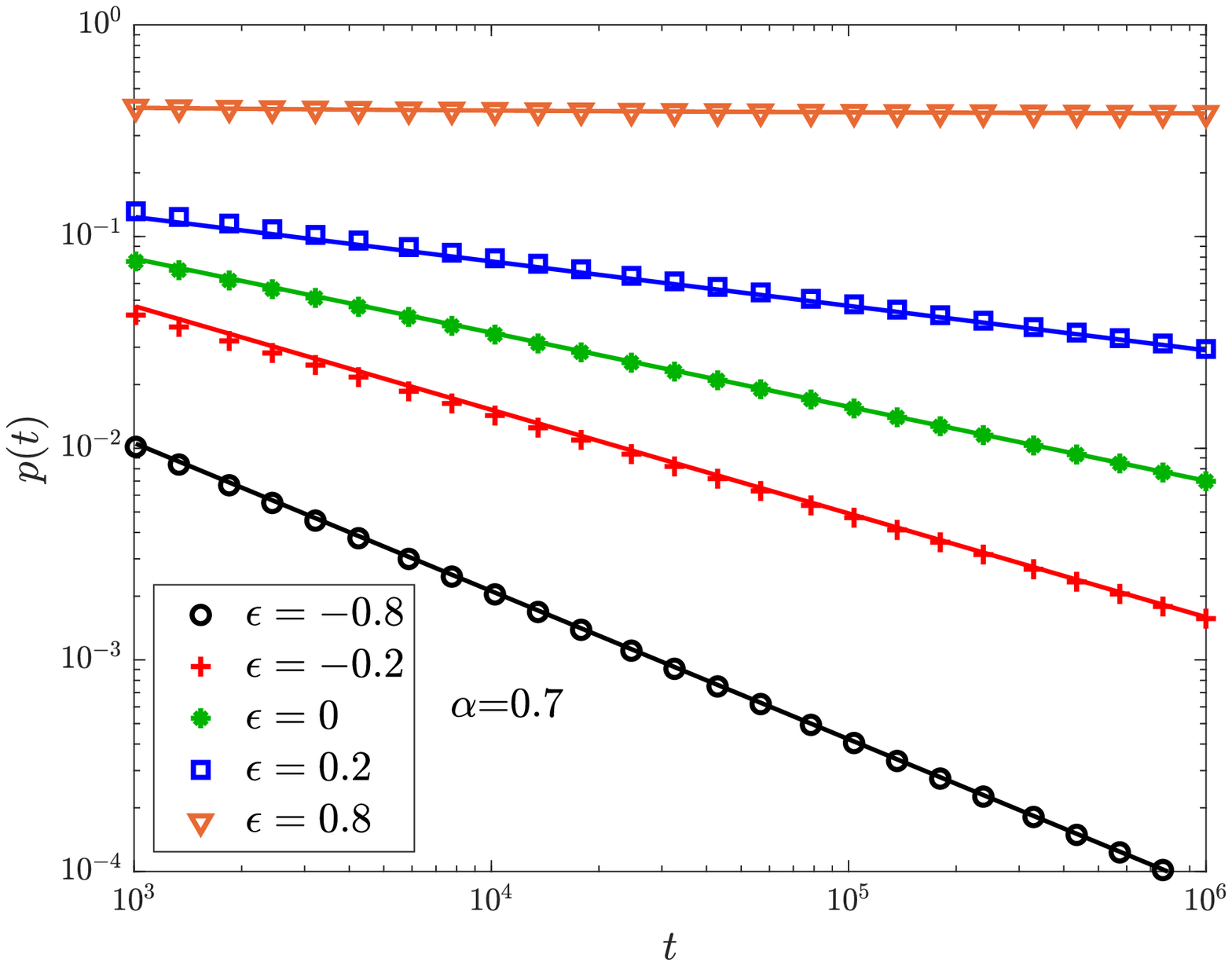}
		\caption{}\label{fig: ret-0}
	\end{subfigure}
	\begin{subfigure}{.45\textwidth}
		\centering
		\includegraphics[width=\linewidth]{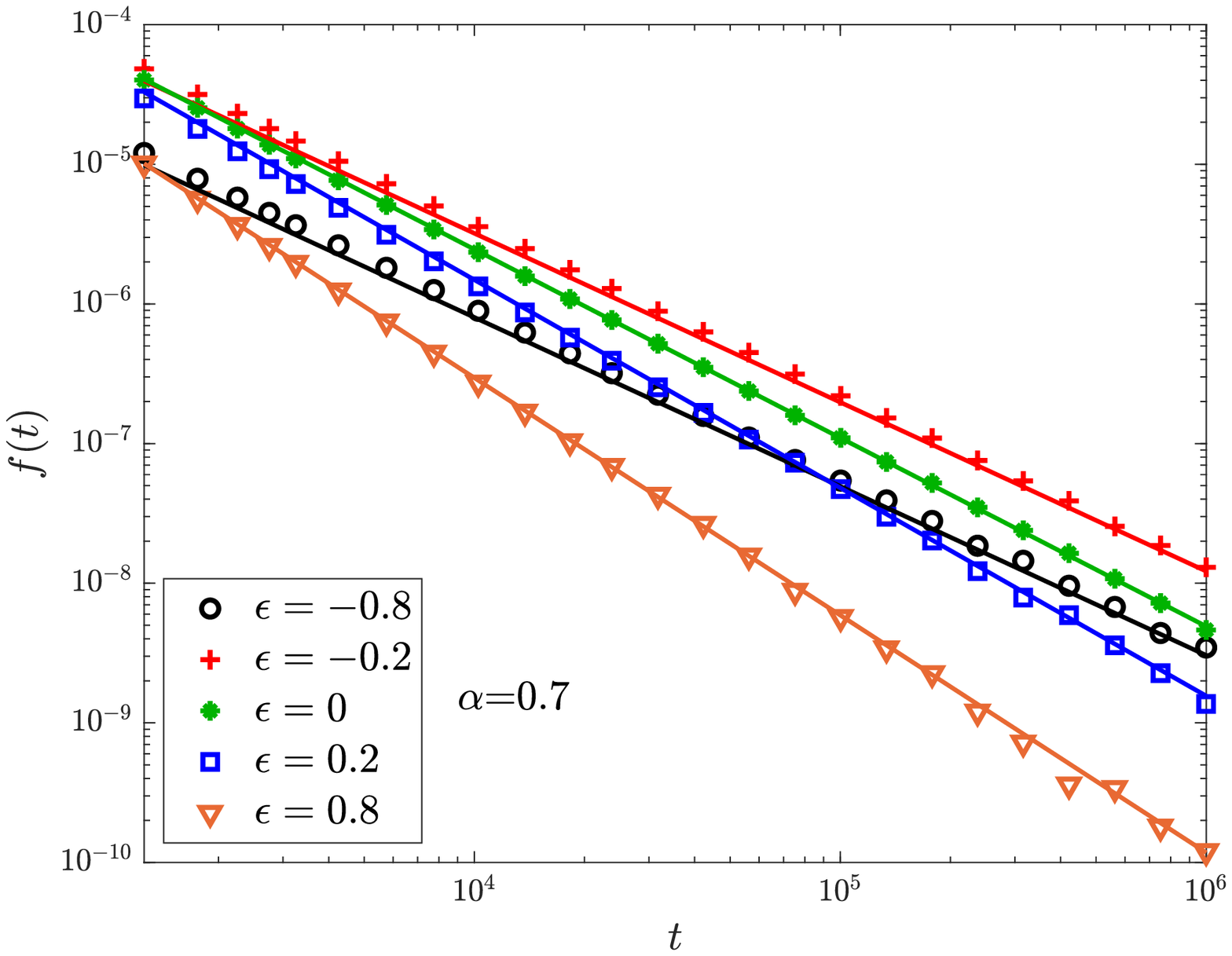}
		\caption{}\label{fig: first-ret-0}
	\end{subfigure}
	\caption{Data are obtained simulating $10^8$ walks up to time $10^6$ with $\alpha=0.7$, $t_0=1$. Markers represent the simulation results for different values of $\epsilon$, lines the corresponding theoretical predictions. We use logarithmic scales on both the horizontal and vertical axes. (\textbf{a}) The PDF $p(t)$ of being at the origin. (\textbf{b}) The first-return PDF $f(t)$.}
\end{figure}

Here we compare Figures~\ref{fig: ret-0} and \ref{fig: first-ret-0} with Eq.~\eqref{asympt-ret-0} and Eq.~\eqref{asympt-first-ret-0}, respectively: there is good agreement with the previous theoretical analysis.

\subsection{Occupation times}
\begin{figure}[H]
	\centering
	\begin{subfigure}{.45\textwidth}
		\centering
		\includegraphics[width=\linewidth]{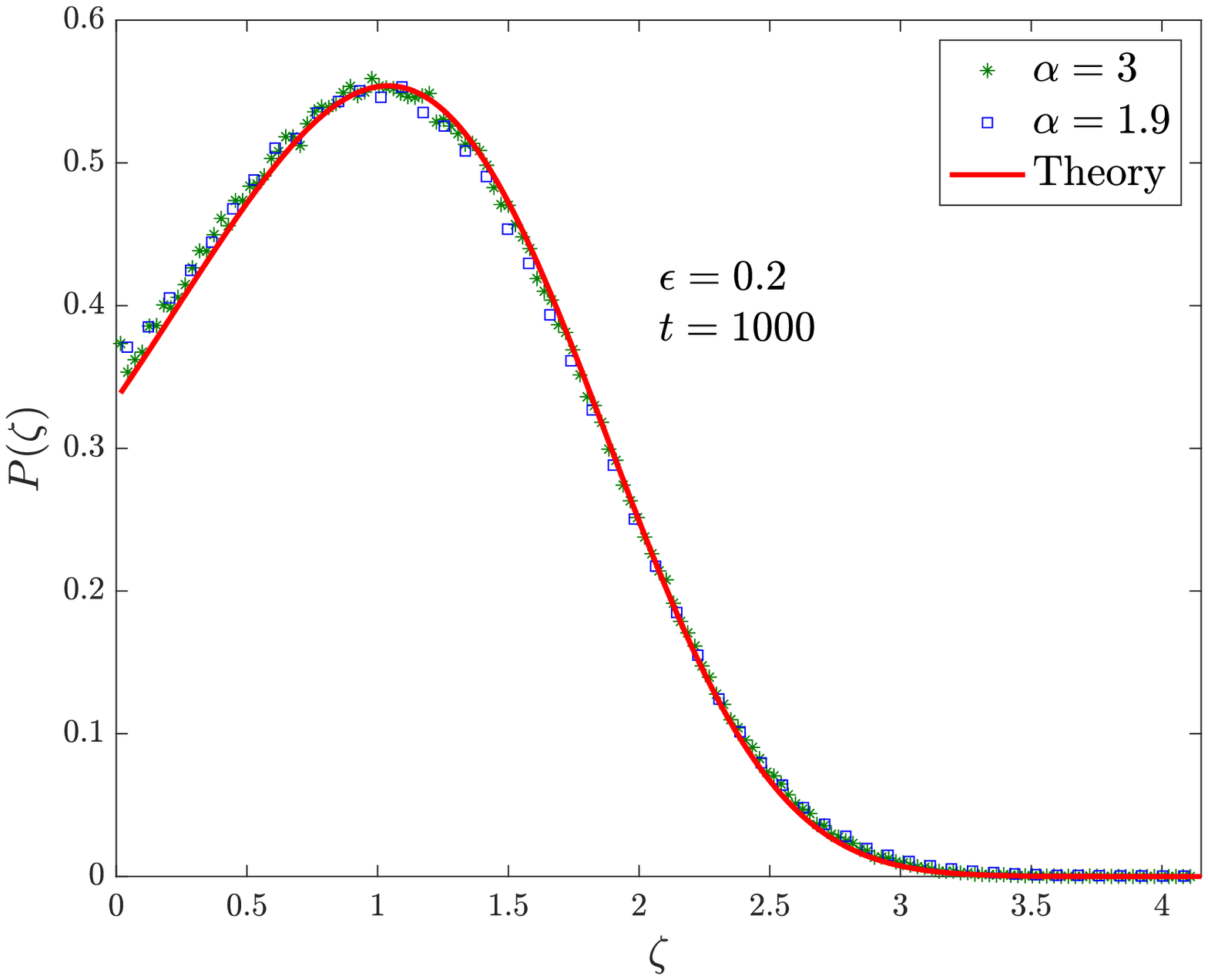}
		\caption{}\label{fig: non-erg-or-fin}
	\end{subfigure}
	\begin{subfigure}{.45\textwidth}
		\centering
		\includegraphics[width=\linewidth]{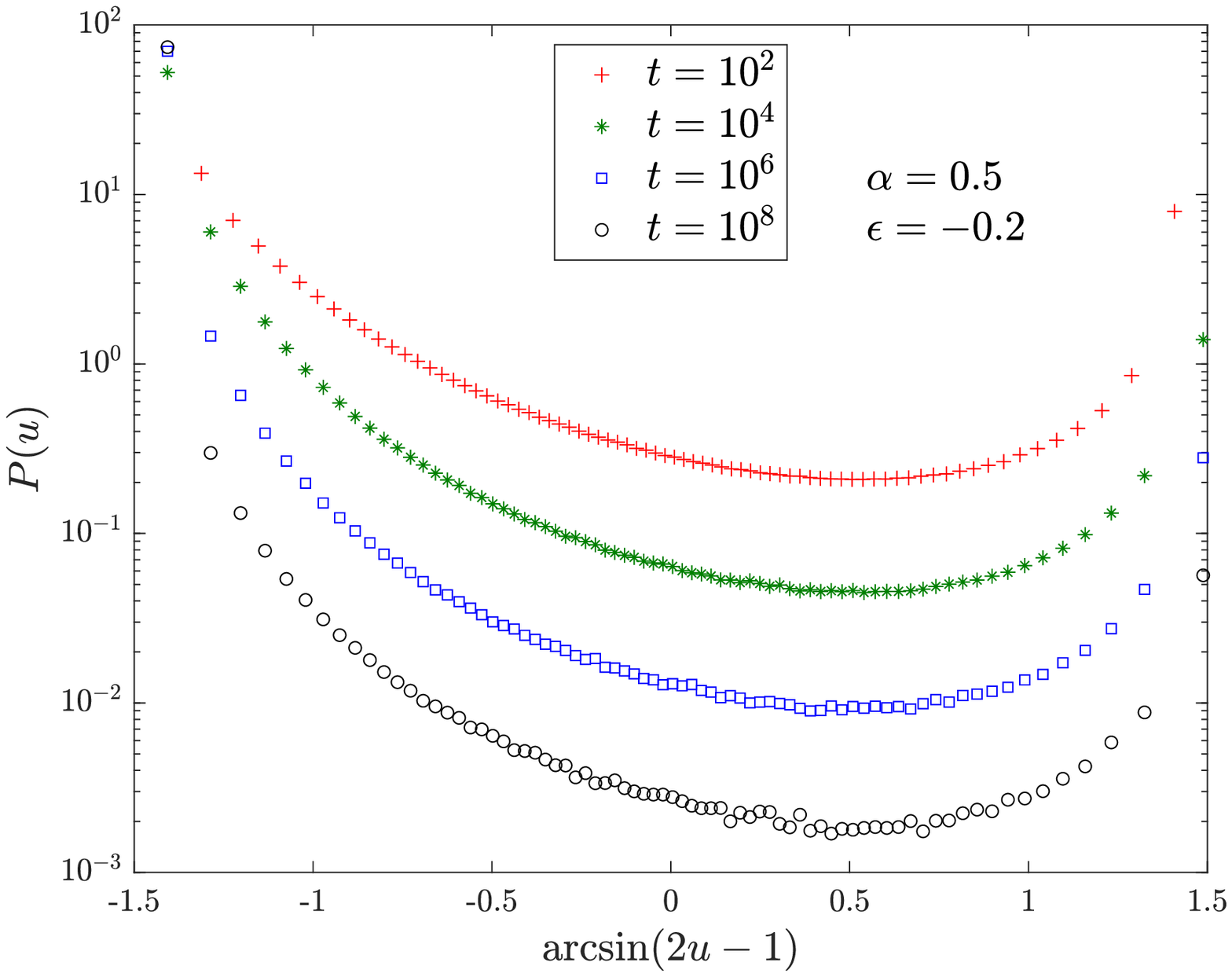}
		\caption{}\label{fig: non-erg-or-inf}
	\end{subfigure}
	\caption{Markers represent the numerical results obtained simulating $10^7$ walks with a cut-off $t_0=0.1$. (\textbf{a}) The PDF $P(\zeta)$ of the rescaled fraction of continuous-time $\zeta\coloneqq L(t/\tau)\tau^{\rho-1} T_{in}(t)/t^{\rho}$ spent at the origin when $\epsilon=0.2$, $\alpha=3$ or $\alpha=1.9$, for an evolution time of $10^3$. (\textbf{b}) The PDF $P(u)$ of the fraction of continuous-time $u\coloneqq T_{in}(t)/t$ spent at the origin for the case $\epsilon=-0.2$, $\alpha=0.5$ evolved up to different times. In addition, we perform the transformation $u\mapsto \arcsin(2u-1)$ on the horizontal axis.}
	\label{fig: non-erg-or}
\end{figure}

In Figure~\ref{fig: non-erg-or} we examine the dependence of the PDF of the occupation time of the origin on the features of the waiting-time distribution in the purely non-ergodic regime $\epsilon \in \left(-\frac 1 2,\frac 1 2\right)$. In the first one, Figure~\ref{fig: non-erg-or-fin}, we have $\alpha > 1$, namely a finite first moment with a finite ($\alpha=3$) or infinite ($\alpha=1.9$) variance, and as a consequence in both cases the occupation time $T_{in}(t)$ rescaled by its mean value, $\zeta$, follows a limiting Mittag-Leffler distribution of index $\rho=\frac 1 2+\epsilon$, which is the same as the properly rescaled number of visits to the origin. In the presence of an infinite first moment, instead, there is no longer an appropriate scaling function: we show (Figure~\ref{fig: non-erg-or-inf}) the slow convergence of the fraction of occupation time $\frac{T_{in}(t)}{t}$ to a Dirac delta with mass at $0$. For increasing evolution times, the peak at $u=0$ becomes more and more prominent with respect to $u=1$ in the asymmetric U-shaped PDF, suggesting the collapse to a degenerate Lamperti distribution.

\begin{figure}[H]
	\centering
	\begin{subfigure}{.45\textwidth}
		\centering
		\includegraphics[width=\linewidth]{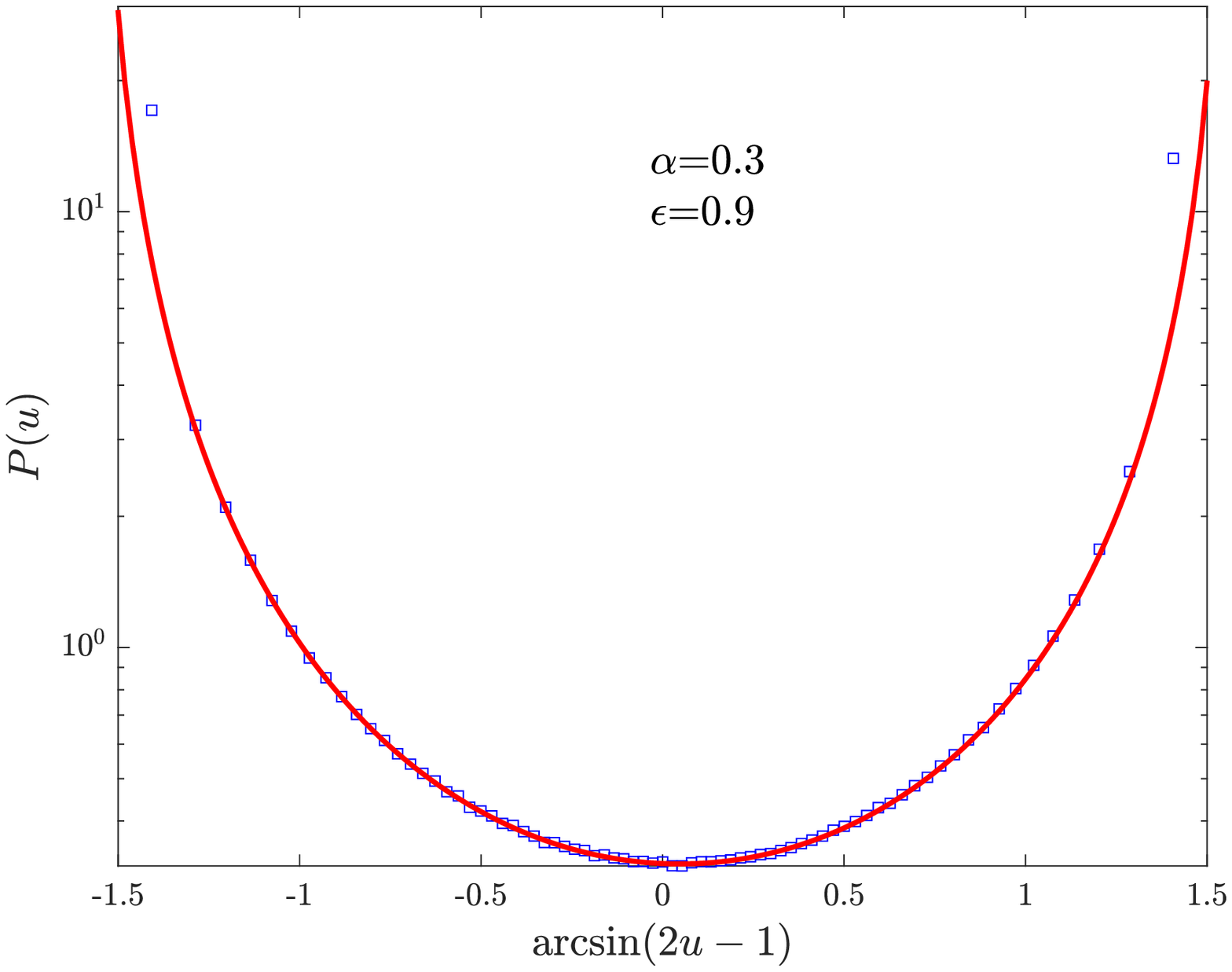}
		\caption{}\label{fig: erg-or-min}
	\end{subfigure}
	\begin{subfigure}{.45\textwidth}
		\centering
		\includegraphics[width=\linewidth]{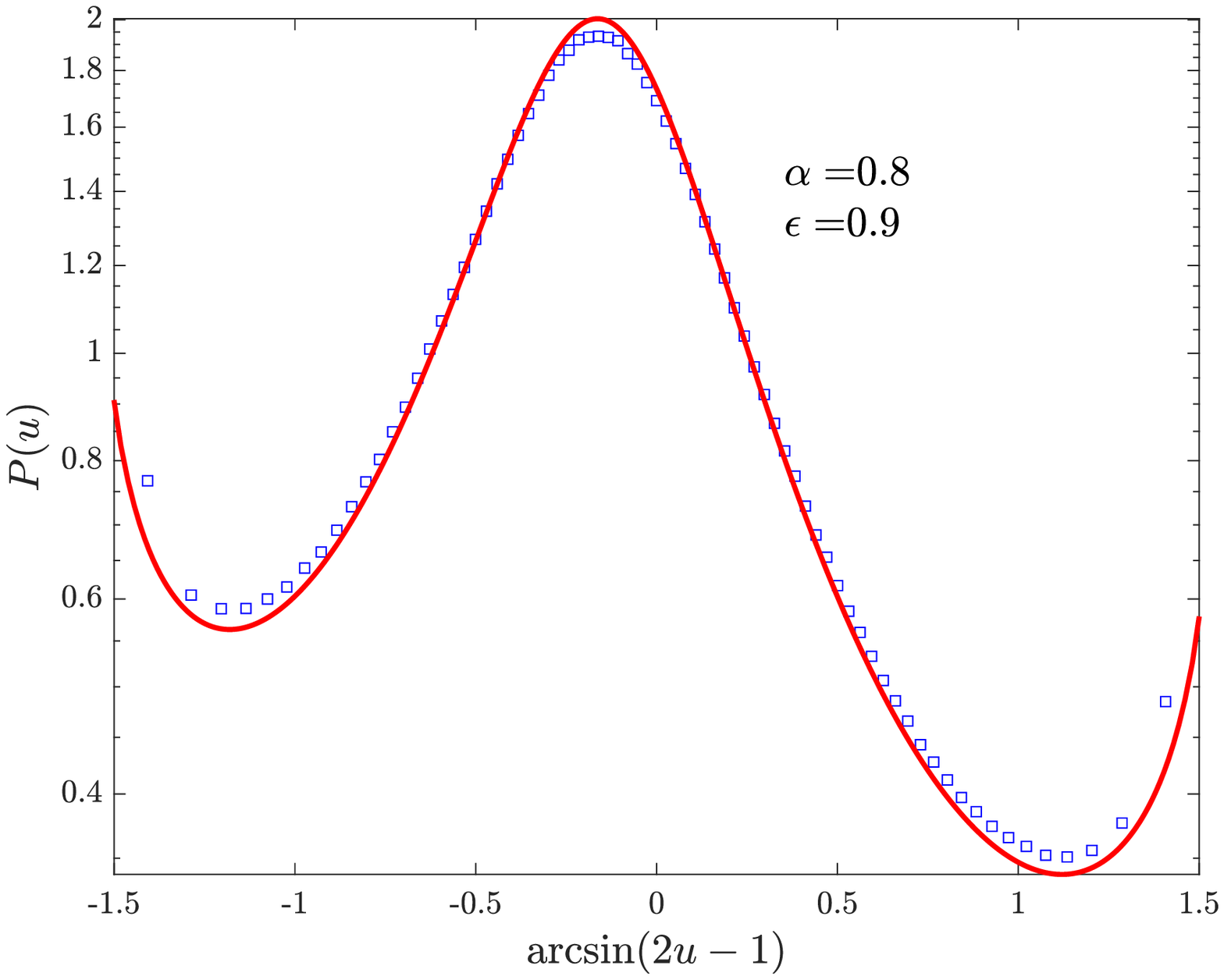}
		\caption{}\label{fig: erg-or-max}
	\end{subfigure}
	\caption{PDF $P(u)$ of the fraction of continuous-time $u\coloneqq T_{in}(t)/t$ spent at the origin. Markers represent the simulation results, (red) lines the theoretical Lamperti distributions. We choose $\epsilon=0.9$ and $t_0=1$. Moreover, for the sake of readability, we perform the transformation $u\mapsto \arcsin(2u-1)$ on the horizontal axis. (\textbf{a}) The case $\alpha=0.3$ obtained simulating $10^7$ walks up to time $10^7$. (\textbf{b}) The case $\alpha=0.8$ with $10^7$ walks evolved up to time $10^8$.}
	\label{fig: erg-or}
\end{figure}

Next, as illustrated in Figure~\ref{fig: erg-or}, we move on to the ergodic regime of the underlying random walk: we consider different values for $\alpha$ in order to hint that, when $\alpha$ approaches $1$, the expected Lamperti distribution, Eq.~\eqref{eq-lamp-occ-time}, eventually collapses to a Dirac delta centered at the mean value $\eta$ of the occupation time, according to previous results in the physical literature \cite{J. Stat. Phys. 3, J. Phys. Condens. Matter 1, Phys. Rev. E 3, Phys. Rev. Lett. 1}.

\begin{figure}[H]
	\centering
	\includegraphics[width=.5\textwidth]{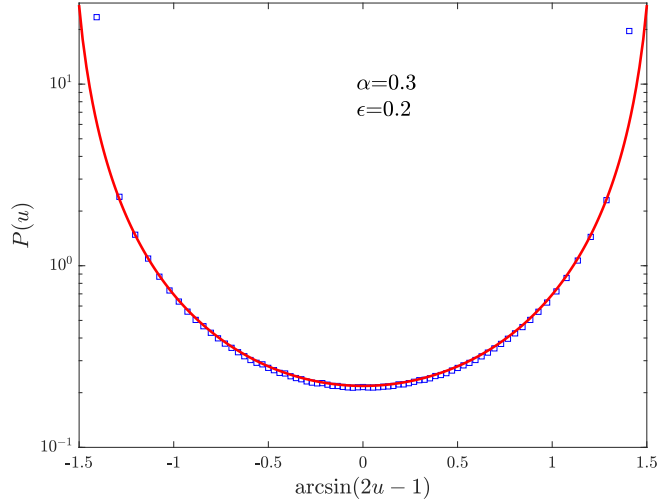}
	\caption{PDF $P(u)$ of the fraction of continuous-time $u\coloneqq T_{out}(t)/(2t)$ spent in the positive semi-axis for $10^7$ walks evolved up to time $10^{11}$ with $\epsilon=0.2$, $\alpha=0.3$ and $t_0=1$. Markers represent numerical data, the (red) line the theoretical Lamperti distribution. Moreover, for the sake of readability, we perform the transformation $u\mapsto \arcsin(2u-1)$ on the horizontal axis.}
	\label{fig: non-erg-pos-ax}
\end{figure}

\begin{figure}[H]
	\centering
	\begin{subfigure}{.45\textwidth}
		\centering
		\includegraphics[width=\linewidth]{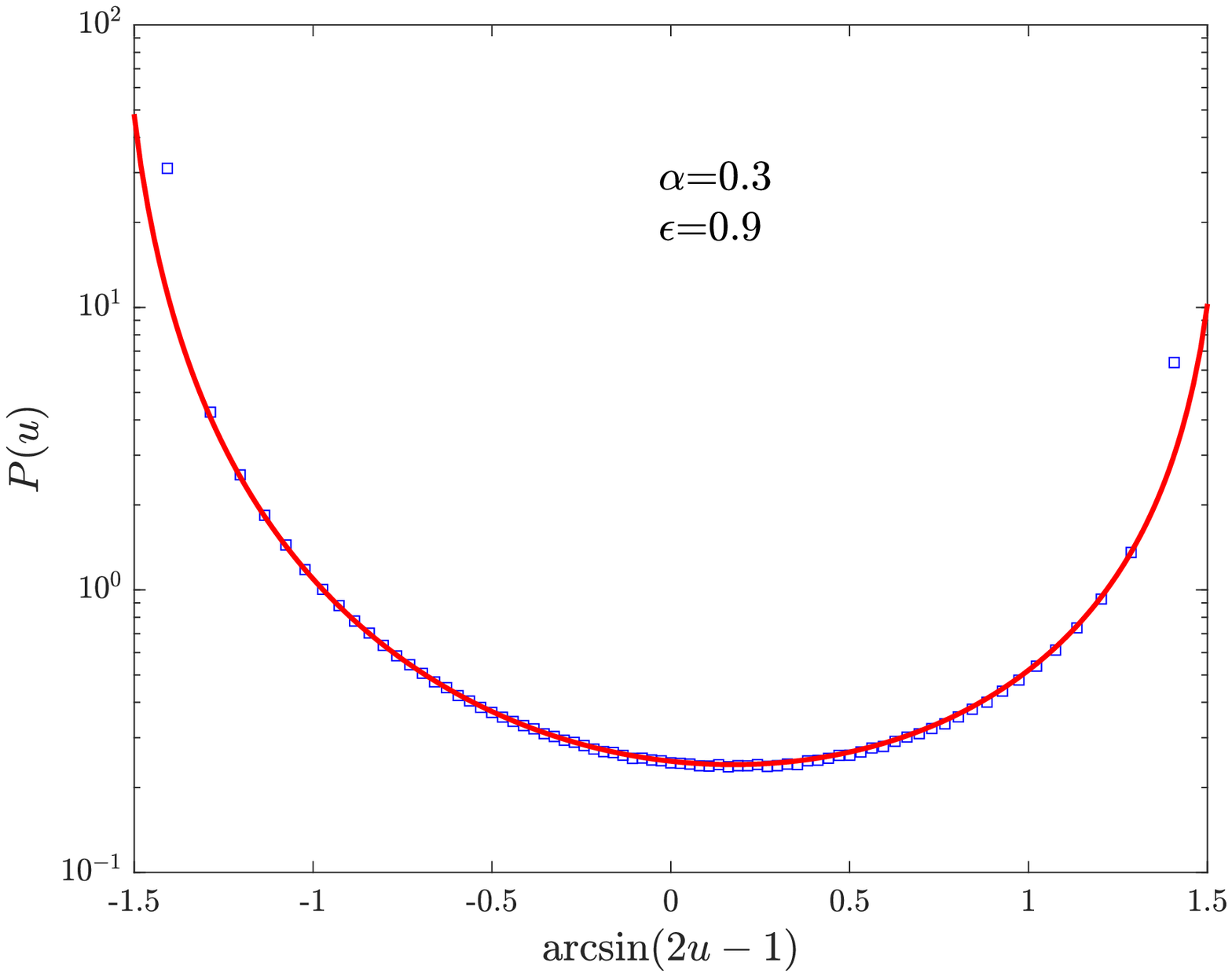}
		\caption{}\label{fig: erg-pos-ax-min}
	\end{subfigure}
	\begin{subfigure}{.45\textwidth}
		\centering
		\includegraphics[width=\linewidth]{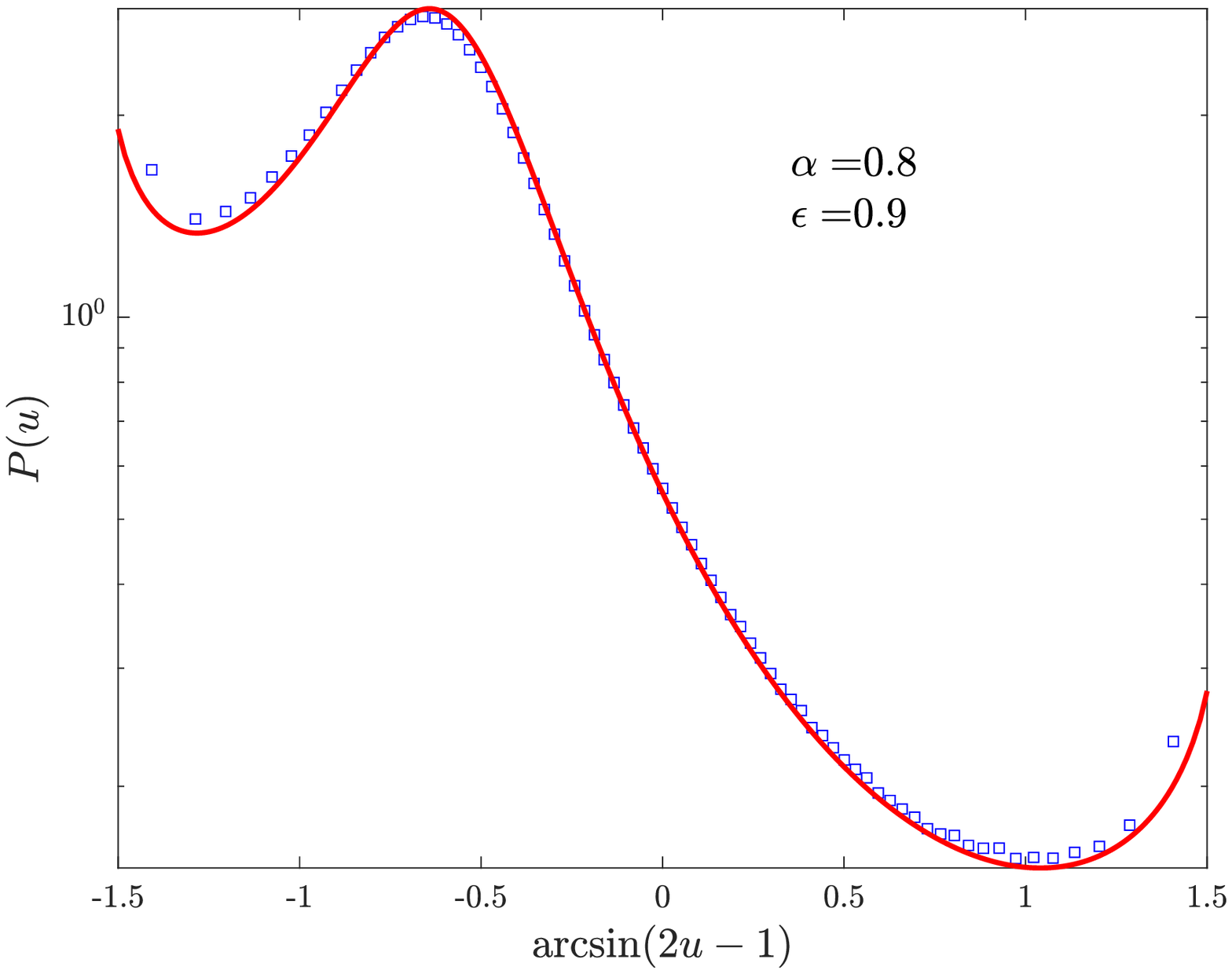}
		\caption{}\label{fig: erg-pos-ax-max}
	\end{subfigure}
	\caption{PDF $P(u)$ of the fraction of continuous-time $u\coloneqq T_{out}(t)/(2t)$ spent in the positive semi-axis. Markers represent the simulation results, (red) lines the theoretical Lamperti distributions. We choose $\epsilon=0.9$ and $t_0=1$. Moreover, for the sake of readability, we perform the transformation $u\mapsto \arcsin(2u-1)$ on the horizontal axis. (\textbf{a}) The case $\alpha=0.3$ obtained simulating $10^7$ walks up to time $10^7$. (\textbf{b}) The case $\alpha=0.8$ with $10^7$ walks evolved up to time $10^8$.}
	\label{fig: erg-pos-ax}
\end{figure}

We now discuss the distribution of the occupation time of the positive semi-axis. In Figure~\ref{fig: non-erg-pos-ax}, we take a purely non-ergodic process: since the fraction of time spent at the origin is negligible, we have the expected symmetric Lamperti distribution of index $\alpha\rho$, which replaces the discrete-time parameter $\rho$. In Figure~\ref{fig: erg-pos-ax}, we shift to the discrete-time ergodic regime by setting $\epsilon=0.9$. We can observe once again the birth of the continuous-time ergodic regime when $\alpha \to 1$, with an asymmetry due to the fact that $\frac{T_{in}(t)}{t}\not\to 0$.

\subsection{Moments spectrum}
\begin{figure}[H]
	\centering
	\begin{subfigure}{.45\textwidth}
		\centering
		\includegraphics[width=\linewidth]{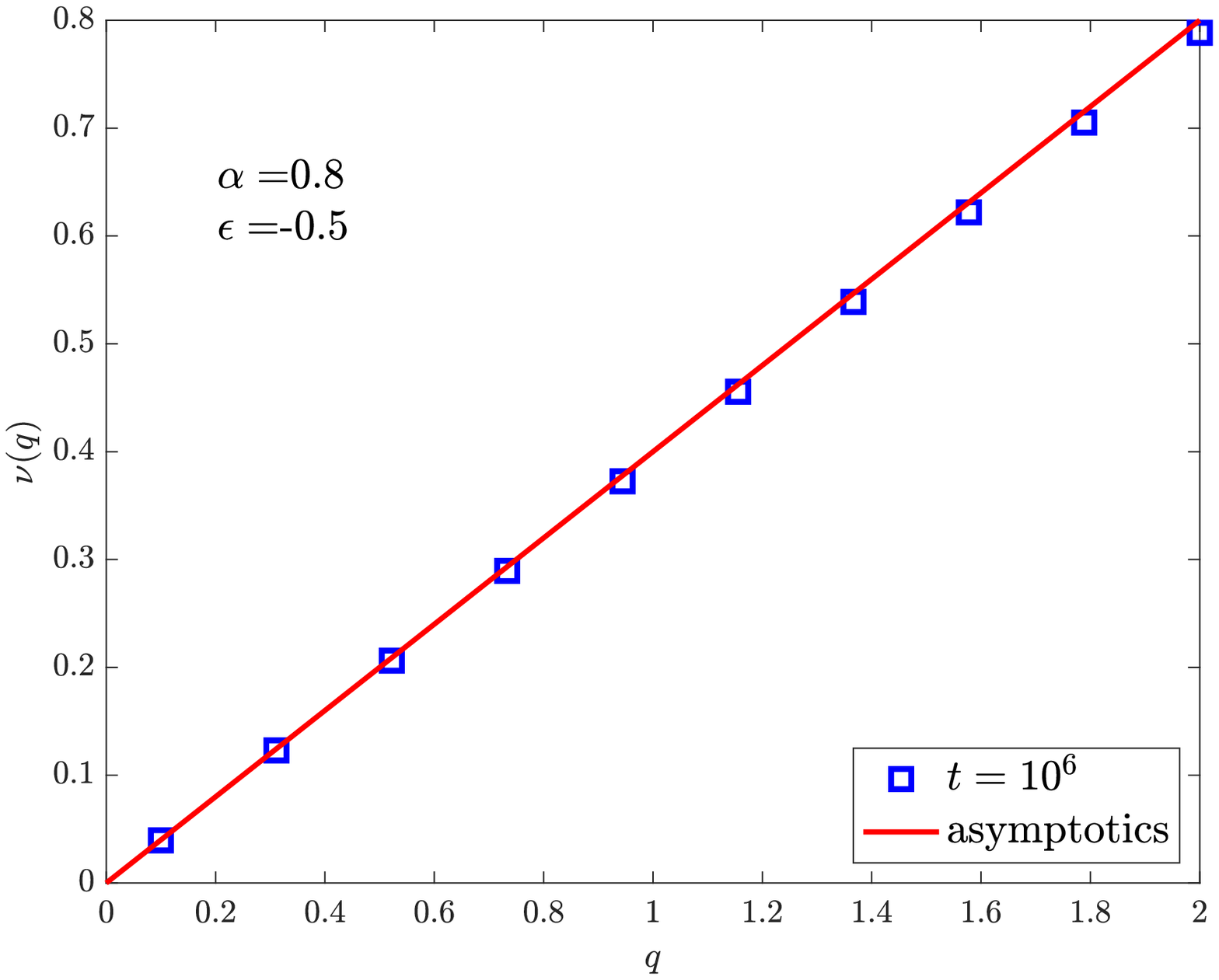}
		\caption{}\label{fig: mom-1}
	\end{subfigure}
	\begin{subfigure}{.45\textwidth}
		\centering
		\includegraphics[width=\linewidth]{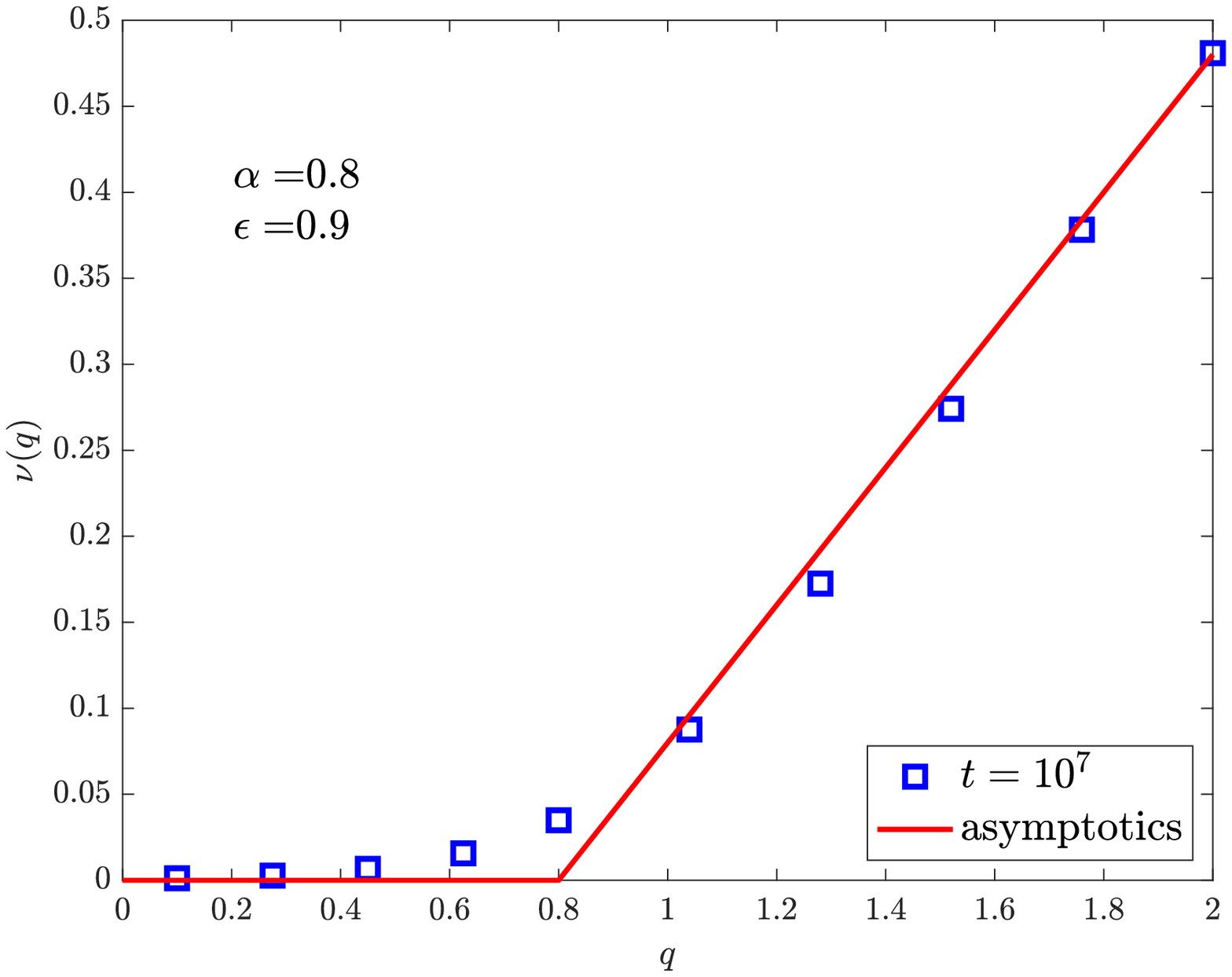}
		\caption{}\label{fig: mom-2}
	\end{subfigure}
	\caption{Comparison between the asymptotic exponents of the power-law growth in time of the $q$-th moments. Markers represent the simulation results, lines the theoretical predictions. $\alpha$ is set equal to $0.8$ and $t_0=1$. (\textbf{a}) The case $\epsilon=-0.5$. The figure is obtained by simulating $10^5$ walks up to time $10^6$. (\textbf{b}) The case $\epsilon=0.9$ with $10^6$ walks evolved up to time $10^7$.}
\end{figure}

In Figure~\ref{fig: mom-1} you can see the expected smooth behaviour for a purely non-ergodic process, although it is no longer related to normal diffusion. In Figure~\ref{fig: mom-2}, instead, in addition to subdiffusion we recognize the presence of a corner, since for $q<2\epsilon-1$ the moments tend to a constant, which is typical of the underlying ergodic property: the convergence near the critical point is slower. 
 
\subsection{Records}
\begin{figure}[H]
	\centering
	\includegraphics[width=.5\textwidth]{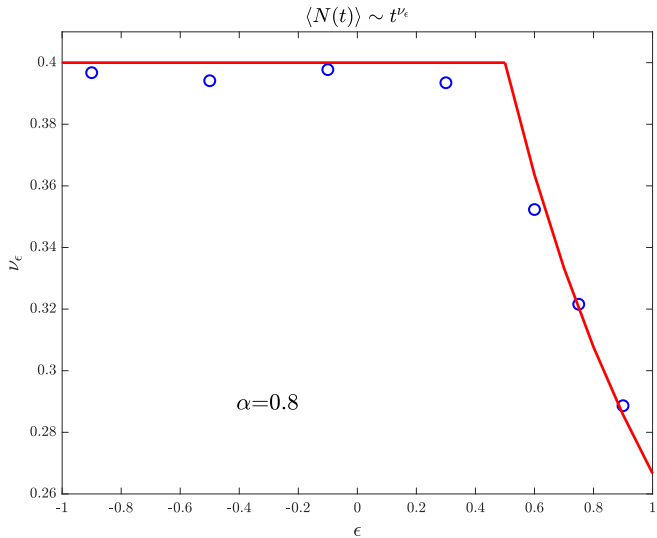}
	\caption{Exponent characterizing the power-law growth of the number of records with respect to the time. We consider $10^6$ walks and an evolution time of $10^7$. Again we have $\alpha=0.8$ and $t_0 =1$. Markers refer to data and the (red) line to the theoretical prediction.}
	\label{fig: rec}
\end{figure}

In Figure~\ref{fig: rec}, we finally show the asymptotic behaviour of the mean number of records, or equivalently the expected maximum, up to time $t$. In particular, we want to emphasize that, even if the range $\epsilon \in \left( -\frac 1 2,\frac 1 2 \right)$ becomes an anomalous regime (in contrast with the discrete-time model), the mean number of records still behaves as the first moment.

\section{Discussion}\label{discussion}
We have reassessed all the exact results found out in our previous work \cite{J. Stat. Mech. 1}  in the light of the continuous-time formalism. By considering waiting times on the sites picked from a heavy-tailed distribution lacking the first moment, meaningful modifications in all regimes can be carried out.

By tuning the real parameter $|\epsilon|<1$, we detect the following differences with respect to the discrete-time dynamics. First of all, the ergodic regime for $\epsilon >\frac 1 2$ fades out. Nevertheless, the underlying ergodic property makes the continuous-time upper range distinct from the purely non-ergodic processes $\left(\epsilon \leq \frac 1 2\right)$: visits at the origin have more and more weight since the fraction of time spent at the starting site does not converge to $0$. Although in the presence of an infinite mean recurrence time, due solely to the irregular temporal component, we have a non-degenerate Lamperti distribution for the quantity of interest. Secondly, the strong-anomalous diffusion regime, characterizing the ergodic processes in the discete-time version, is \textit{weakly} extended to the purely non-ergodic range, where \textit{weak} subdiffusion replaces normal diffusion. More generally, return and first-return probabilities have a slower asymptotic power-law decay, depending on the parameter $\alpha$ of the temporal tail bounds.

We hope our studies will fall under an increasingly wide class of general exact results for stochastic processes lacking translational invariance, which hide subtle phenomena of physical interest not satisfied by the well-known homogeneous counterpart.

\vspace{1cm} \noindent 
{{\bf Author contributions:} All authors have contributed substantially to the work. All authors have read and agreed to the published version of the manuscript.}

\vspace*{0.5 cm}\noindent 
{{\bf Funding:} The authors acknowledge partial support from PRIN Research Project No. 2017S35EHN "Regular and stochastic behavior in dynamical systems" of the Italian Ministry of Education, University and Research (MIUR).}

\vspace*{0.5 cm}\noindent 
{{\bf Conflicts of Interest:} The authors declare no conflict of interest. The funders had no role in the design of the study; in the collection, analyses, or interpretation of data; in the writing of the manuscript, or in the decision to publish the results.} 

\vspace*{1 cm}\noindent 
{\bf Abbreviations: }The following abbreviations are used in this manuscript:\\

\noindent 
\begin{tabular}{@{}ll}
CTRW & Continuous Time Random Walk\\
PDF & Probability Density Function\\
i.i.d. & Independent Identically Distributed
\end{tabular}


\appendix
\section{Gillis-type proof}\label{app-gillis-way}
In Section~\ref{sec-gillis-way} we have written the following equation:

\begin{equation}
p^a(j,t)-\delta_{j,0}\delta(t)=\int_0^t p^a(j-1,t')\psi(t-t')dt'\mathcal{R}_{j-1}+\int_0^tp^a(j+1,t')\psi(t-t')dt' \mathcal{L}_{j+1}.
\end{equation}

In the Laplace domain it reads:

\begin{equation}
\hat{p}^a(j;\: s)-\delta_{j,0}=\hat{\psi}(s)\left[\hat{p}^a(j-1;\: s)R_{j-1}+\hat{p}^a(j+1;\:s)L_{j+1}\right],
\end{equation}
and considering also the generating function on sites, we get:

\begin{equation}
\hat{P}^a(x,s)-1=\hat{\psi}(s)\left[x\hat{R}(x,s)+\frac{1}{x}\hat{L}(x,s)\right]\qquad \mbox{where}\qquad \hat{R}(x,s):=\sum_{j=-\infty}^{+\infty}\hat{p}^a(j;\:s)\mathcal{R}_jz^j,
\end{equation}
similarly for $\hat{L}(x,s)$ and clearly $\hat{R}(x,s)+\hat{L}(x,s)=\hat{P}^a(x,s)$. As a consequence, we can write:

\begin{equation}
\frac 1 2 \hat{P}^a(x,s)-\frac{\epsilon}{2}\sum_{j\neq 0}\frac{\hat{p}^a(j;s)}{j}x^j=\hat{R}(x,s)=
\frac{x-\hat{\psi}(s)}{\hat{\psi}(s)(x^2-1)}\hat{P}^a(x,s)-\frac{x}{\hat{\psi}(s)(x^2-1)}.
\end{equation}

Differentiating both sides with respect to $x$, we obtain:

\begin{equation}
\frac{\partial}{\partial x}\hat{R}(x,s)=\frac 1 2 \frac{\partial}{\partial x}\hat{P}^a(x,s)-\frac 1 2 \frac{\epsilon}{x}[\hat{P}^a(x,s)-\hat{p}^a(0;\:s)],
\end{equation}
to be compared with:

\begin{equation}
\hat{\psi}(s)(x^2-1)^2\frac{\partial}{\partial x}\hat{R}(x,s)=\hat{P}^a(x,s)[2x\hat{\psi}(s)-(x^2+1)]+\frac{\partial}{\partial x}\hat{P}^a(x,s)(x^2-1)(x-\hat{\psi}(s))+x^2+1,
\end{equation}
and so the differential equation for $\hat{P}^a(x,s)$ is:

\begin{multline}
[x\hat{\psi}(s)(x^2-1)^2-2x(x^2-1)(x-\hat{\psi}(s))]\frac{\partial}{\partial x} \hat{P}^a(x,s)+[2x(x^2+1)-4x^2\hat{\psi}(s)-\epsilon \hat{\psi}(s)(x^2-1)^2]\hat{P}^a(x,s)\\
=2x(x^2+1)-\epsilon \hat{\psi}(s)(x^2-1)^2\hat{p}^a(0;\:s).
\end{multline}

Now, we set $x=e^{i\phi}$, $\frac{\partial}{\partial x} \hat{P}^a(x,s)\eqqcolon\frac{\partial}{\partial x}E(\phi(x),s)=-\frac{i}{x}\frac{\partial E}{\partial \phi}$ and split real and imaginary parts, thus we get $\frac{\partial E}{\partial \phi}+f(\phi)E=g(\phi)$, where:

\begin{eqnarray}
f(\phi) &=& [1-\hat{\psi}(s)\cos\phi]^{-1}[\hat{\psi}(s)(1-\epsilon)\sin\phi-(1-\hat{\psi}(s)\cos\phi)\cot\phi],\\
g(\phi) &=& [1-\hat{\psi}(s)\cos\phi]^{-1}[-\cot\phi-\epsilon\hat{\psi}(s)\sin\phi\hat{p}^a(0;\:s)],
\end{eqnarray}
and the solution is:

\begin{equation}
E(\phi,s)=e^{-\int f(\phi)d\phi}\left[\int g(\phi)e^{\int f(\phi)d\phi}d\phi+ \mbox{const.} \right].
\end{equation}

In order to recover Eq.~\eqref{gillis-res}, it is sufficient to perform the calculations and recall that:

\begin{equation}
\hat{p}^a(0;\: s)=\frac{1}{2\pi}\int_0^{2\pi}E(\phi,s)d\phi.
\end{equation}
\section{Hitting time PDF of the origin: exact results}\label{app-exact-ret}
Let us derive the exact and asymptotic behaviours of the probabilities of being at the origin. For the moment, we neglect the limiting cases $\epsilon=\pm \frac 1 2$. All we need are the following properties of the Gamma function and the transformation formula $15.3.6$ for the hypergeometric functions in \cite{Handbook 1}:

\begin{equation}
\Gamma(z+1)=z\Gamma(z),\quad z\notin \mathbbm{Z}_-, \qquad\qquad \prod_{k=0}^{n-1}\Gamma\left(z+\frac{k}{n}\right)=(2\pi)^{\frac{n-1}{2}}n^{\frac 1 2-nz}\Gamma(nz),
\end{equation}

\begin{eqnarray}
_{2}F_1(a,b;\:c;\:z)&=& \frac{\Gamma(c)\Gamma(c-a-b)}{\Gamma(c-a)\Gamma(c-b)} {}_{2}F_1(a,b;\:a+b-c+1;\:1-z)\\
&&\quad +(1-z)^{c-a-b}\frac{\Gamma(c)\Gamma(a+b-c)}{\Gamma(a)\Gamma(b)} {}_{2}F_1(c-a,c-b;\:c-a-b+1;\:1-z)\\
&\eqqcolon&G\cdot F_G + (1-z)^{c-a-b} K \cdot F_K,
\end{eqnarray}
where $|\arg(1-z)|<\pi$ and $c-a-b\notin\mathbbm{Z}$. Now, recalling Eq.~\eqref{pr-hitting-laplace} and considering $\hat{\psi}(s)\sim 1-bs^{\alpha}$ for $s\to 0$, we have to asymptotically compute $\hat{p}(j_0,0;\: s)$ in different regimes.

First of all, let us take $\epsilon \in \left(-1,-\frac 1 2\right)$. By skipping the intermediate steps, we get:

\begin{equation}
\hat{p}(j_0,0;\: s) \sim \frac{b}{s^{1-\alpha}}\left(\frac 1 2\right)^{|j_0|}\frac{\Gamma(1+\epsilon+|j_0|)}{|j_0|!\Gamma(1+\epsilon)}\frac{G_N}{G_D}=\frac{b}{s^{1-\alpha}}\frac{\Gamma(1+\epsilon+|j_0|)}{2^{|j_0|}|j_0|!\Gamma(1+\epsilon)}\frac{2^{|j_0|}\Gamma(|j_0|+1)\Gamma(1-\epsilon)}{(-2\epsilon-1)\Gamma(|j_0|-\epsilon)},
\end{equation}
since:

\begin{equation}
0<c_N-a_N-b_N=-\epsilon-\frac 1 2<\frac 1 2 \qquad \mbox{and}\qquad 0<\frac 1 2 < c_D-a_D-b_D=\frac 1 2 -\epsilon<\frac 3 2.
\end{equation}

In conclusion, by means of Tauberian theorems:

\begin{equation}
p(j_0,0,t)\sim \frac{b}{-2\epsilon-1}\frac{\Gamma(1+\epsilon+|j_0|)\Gamma(1-\epsilon)}{\Gamma(1+\epsilon)\Gamma(|j_0|-\epsilon)}\frac{1}{\Gamma(1-\alpha)}\frac{1}{t^{\alpha}}\qquad \mbox{as}\quad t \to \infty.
\end{equation}

Before going any further, let us make a comment which highlights the transience property. It is glaring that:

\begin{equation}
\hat{p}(j_0,0;\: s)\propto \frac{1-\hat{\psi}(s)}{s}=\hat{\chi}_0(s),
\end{equation}
namely it has the same scaling law of the survival probability on a site, although possibly not the same coefficient. In particular, setting $j_0=0$ means to consider return probabilities. But if you choose moreover the limiting case $\epsilon=-1$, when the particle moves to $\pm 1$  it will never come back since $\mathcal{L}_{+1}=0=\mathcal{R}_{-1}$ and it is like placing two barriers with total reflection on the outside. As a consequence $p(t)\equiv \chi_0(t)$.

If $\epsilon \in \left(-\frac 1 2,\frac 1 2\right)$, we find again $0<c_D-a_D-b_D(<1)$ but $(-1<)c_N-a_N-b_N<0$ and so when $s\to 0$ and $t\to \infty$:

\begin{eqnarray}
\hat{p}(j_0,0;\: s) &=& \frac{1-\hat{\psi}(s)}{s}\left(\frac{\hat{\psi}(s)}{2}\right)^{|j_0|}\frac{\Gamma(1+\epsilon+|j_0|)}{|j_0|!\Gamma(1+\epsilon)}[1-\hat{\psi}^2(s)]^{-\frac 1 2 -\epsilon}\frac{K_NF_{K_N}+[1-\hat{\psi}^2(s)]^{\frac 1 2+\epsilon}G_NF_{G_N}}{G_DF_{G_D}+[1-\hat{\psi}^2(s)]^{\frac 1 2-\epsilon}K_DF_{K_D}}\nonumber \\
&\sim& \frac{b}{s^{1-\alpha}}\frac{\Gamma(1+\epsilon+|j_0|)}{2^{|j_0|}|j_0|!\Gamma(1+\epsilon)}(2bs^{\alpha})^{-\epsilon-\frac 1 2}\frac{K_N }{G_D }\nonumber\\
& =&\frac{1}{2^{\epsilon+\frac 1 2 }}\frac{b^{\frac 1 2 -\epsilon}}{s^{1-\frac{\alpha}{2}+\alpha\epsilon}}\frac{\Gamma(1+\epsilon+|j_0|)}{2^{|j_0|}|j_0|!\Gamma(1+\epsilon)}2^{2\epsilon+|j_0|}\frac{\Gamma(|j_0|+1)\Gamma\left(\frac 1 2 +\epsilon \right)\Gamma(1-\epsilon)}{\Gamma(\epsilon+1+|j_0|)\Gamma\left(\frac 1 2-\epsilon\right)} ,\\
p(j_0,0,t) &\sim& \left(\frac{b}{2}\right)^{\frac 1 2-\epsilon} \frac{\Gamma\left(\frac 1 2 +\epsilon \right)\Gamma(1-\epsilon)}{\Gamma(1+\epsilon)\Gamma\left(\frac 1 2-\epsilon\right)}\frac{1}{\Gamma\left(1-\frac{\alpha}{2}+\alpha \epsilon\right)}\frac{1}{t^{\alpha\left(\frac 1 2 -\epsilon\right)}}.
\end{eqnarray}

In the first line of $\hat{p}(j_0,0;\: s)$, we want to emphasize that we can always explicitly write the exact slowly-varying function (the last factor), even if then we focus on its asymptotic expansion.

When $\epsilon \in\left(\frac 1 2,1\right)$, instead we have $\left(-\frac 3 2 <\right) c_N-a_N-b_N<-1<0,\:\left(-\frac 1 2 < \right)c_D-a_D-b_D<0$ and as a consequence:

\begin{eqnarray}
\hat{p}(j_0,0;\: s) &\sim& \frac{b}{s^{1-\alpha}}\frac{\Gamma(1+\epsilon+|j_0|)}{2^{|j_0|}|j_0|!\Gamma(1+\epsilon)}\frac{(2bs^{\alpha})^{-\epsilon-\frac 1 2}}{(2bs^{\alpha})^{\frac 1 2-\epsilon}}\frac{K_N }{K_D}\nonumber\\
&=& \frac{1}{2s}\frac{\Gamma(1+\epsilon+|j_0|)}{2^{|j_0|}|j_0|!\Gamma(1+\epsilon)}\frac{2^{1+|j_0|}\Gamma(|j_0|+1)\left(\epsilon-\frac 1 2\right)\Gamma(\epsilon)}{\Gamma(\epsilon+1+|j_0|)}=\frac{1}{s}\frac{2\epsilon-1}{2\epsilon},\\
p(j_0,0,t) &\sim& \frac{2\epsilon-1}{2\epsilon}.
\end{eqnarray}

Finally, we have to handle with the transition points. If $\epsilon=-\frac 1 2$ we need to introduce also $15.3.10$ and $15.3.11$ in \cite{Handbook 1}, with $m=1,2,3,\dots$, $|\arg(1-z)|<\pi$, $|1-z|<1$:

\begin{equation}
_{2}F_{1}(a,b;\: a+b;\: z)=\frac{\Gamma(a+b)}{\Gamma(a)\Gamma(b)}\sum_{n=0}^{\infty}\frac{(a)_n(b)_n}{n!^2}[2\Psi(n+1)-\Psi(a+n)-\Psi(b+n)-\ln(1-z)](1-z)^n,
\end{equation}
where $\Psi(z)=\frac{d}{dz}\ln\Gamma(z)$ denotes the digamma function and $(z)_n$ is the Pochhammer symbol, and:

\begin{multline}
_{2}F_1(a,b;\:a+b+m;\:z)=\frac{\Gamma(m)\Gamma(a+b+m)}{\Gamma(a+m)\Gamma(b+m)}\sum_{n=0}^{m-1}\frac{(a)_n(b)_n}{n!(1-m)_n}(1-z)^n-\frac{\Gamma(a+b+m)}{\Gamma(a)\Gamma(b)}(z-1)^m\\
\times\sum_{n=0}^{\infty}\frac{(a+m)_n(b+m)_n}{n!(n+m)!} (1-z)^n[\ln(1-z)-\Psi(n+1)\\
-\Psi(n+m+1)+\Psi(a+n+m)+\Psi(b+n+m)].
\end{multline}

Hence we get:

\begin{eqnarray}
\hat{p}(j_0,0;\:s) &=& \frac{1-\hat{\psi}(s)}{s}\left(\frac{\hat{\psi}(s)}{2}\right)^{|j_0|}\frac{\Gamma\left(\frac1 2+|j_0|\right)}{|j_0|!\sqrt{\pi}}\frac{_2F_1(a_N,b_N;\:a_N+b_N;\:\hat{\psi}^2(s))}{_2F_1(a_D,b_D;\:a_D+b_D+1;\:\hat{\psi}^2(s))}\\
& \sim &\frac{b}{s^{1-\alpha}}\frac{\Gamma\left(\frac1 2+|j_0|\right)}{2^{|j_0|}|j_0|!\sqrt{\pi}}\frac{\Gamma\left( \frac 3 4 \right) \Gamma\left(\frac 5 4 \right)\Gamma(|j_0|+1)}{\Gamma\left(\frac 1 4+\frac{|j_0|}{2}\right)\Gamma\left(\frac{|j_0|}{2}+\frac 3 4 \right)}\ln\left( \frac{1}{2bs^{\alpha}}\right)=\frac b 4\ln\left( \frac{1}{2bs^{\alpha}}\right)\frac{1}{s^{1-\alpha}},\\
p(j_0,0,t) &\sim & \frac 1 4\ln\left( \frac{t^{\alpha}}{2b}\right)\frac{b}{\Gamma(1-\alpha)}\frac{1}{t^{\alpha}}.
\end{eqnarray}

Whereas, when $\epsilon=+\frac 1 2$, we need also equation $15.3.12$ in \cite{Handbook 1}:

\begin{multline}
_2F_1(a,b;\:a+b-m;\:z)=\frac{\Gamma(m)\Gamma(a+b-m)}{\Gamma(a)\Gamma(b)}(1-z)^{-m}\sum_{n=0}^{m-1}\frac{(a-m)_n(b-m)_n}{n!(1-m)_n}(1-z)^n\\
-\frac{(-1)^m\Gamma(a+b-m)}{\Gamma(a-m)\Gamma(b-m)} \sum_{n=0}^{\infty}\frac{(a)_n(b)_n}{n!(n+m)!}(1-z)^n[\ln(1-z)-\Psi(n+1)\\
-\Psi(n+m+1)+\Psi(a+n)+\Psi(b+n)].
\end{multline}

In this way, we can conclude that:

\begin{eqnarray}
\hat{p}(j_0,0;\:s) &=& \frac{1-\hat{\psi}(s)}{s}\left(\frac{\hat{\psi}(s)}{2}\right)^{|j_0|}\frac{\Gamma\left(\frac 3 2+|j_0|\right)}{|j_0|!\Gamma\left( \frac 3 2 \right)}\frac{_2F_1(a_N,b_N;\:a_N+b_N-1;\:\hat{\psi}^2(s))}{_2F_1(a_D,b_D;\:a_D+b_D;\:\hat{\psi}^2(s))}\\
& \sim &\frac{b}{s^{1-\alpha}}\frac{\Gamma\left(\frac 3 2+|j_0|\right)}{2^{|j_0|}|j_0|!\frac{\sqrt{\pi}}{2}} \frac{\Gamma(|j_0|+1)\Gamma\left(\frac 1 4 \right) \Gamma\left( \frac 3 4 \right)}{\Gamma \left(\frac 3 4 +\frac{|j_0|}{2}\right) \Gamma \left(\frac 5 4+\frac{|j_0|}{2}\right)}\frac{1}{2bs^{\alpha}\ln\left(\frac{1}{2bs^{\alpha}}\right)}\\
&=& \frac{2}{s}\left[\ln\left(\frac{1}{2bs^{\alpha}}\right)\right]^{-1},\\
p(j_0,0,t) &\sim & 2 \left[\ln\left(\frac{t^{\alpha}}{2b}\right)\right]^{-1}.
\end{eqnarray}

\section{First-hitting time PDF: exact results}\label{app-exact-first-ret}
Now, by means of Eq.~\eqref{eq-f-r-0-CT} and Eq.~\eqref{eq-f-r-CT}, we can exploit the previous appendix in order to extend the analysis to first-passage events. Once again, we can deduce exact results although in the end we pull up asymptotic formulas, but in addition this time we have to split our investigation according to the choice of the starting site: we have to differentiate between first-passage and first-return.

\subsection{First-return}\label{subapp-first-ret}
Knowing already that:

\begin{equation}
\hat{p}(s)\sim \begin{cases}
\frac{\epsilon}{2\epsilon+1}\frac{b}{s^{1-\alpha}}\qquad &\mbox{if}\quad -1<\epsilon<-\frac 1 2,\\
\frac 1 4 \ln \left( \frac{1}{2bs^{\alpha}}\right)\frac{b}{s^{1-\alpha}}\qquad &\mbox{if}\quad \epsilon = -\frac 1 2 ,\\
\left(\frac{b}{2}\right)^{\frac1 2 -\epsilon} \frac{\Gamma(1-\epsilon)\Gamma\left(\frac 1 2+\epsilon\right)}{\Gamma\left(\frac 1 2-\epsilon\right)\Gamma(1+\epsilon)}\frac{1}{s^{1-\alpha\left(\frac 1 2-\epsilon\right)}}\qquad & \mbox{if}\quad -\frac 1 2 <\epsilon<+\frac 1 2,\\
2\left[\ln\left(\frac{1}{2bs^{\alpha}}\right)\right]^{-1}\frac 1 s \qquad & \mbox{if}\quad\epsilon=+\frac 1 2,\\
\left(1-\frac{1}{2\epsilon}\right) \frac 1 s\qquad & \mbox{if}\quad+\frac 1 2 <\epsilon<+1,
\end{cases}
\end{equation}
we immediately get:

\begin{equation}
\hat{f}(s)=1-\frac{1-\hat{\psi}(s)}{s\hat{p}(s)}\sim \begin{cases}
1-4 \left[\ln \left( \frac{1}{2bs^{\alpha}}\right)\right]^{-1}\qquad &\mbox{if}\quad \epsilon = -\frac 1 2 ,\\
1-2^{\frac 1 2-\epsilon} \frac{\Gamma\left(\frac 1 2-\epsilon \right)\Gamma(1+\epsilon)}{\Gamma(1-\epsilon)\Gamma\left(\frac1 2+\epsilon\right)}b^{\frac 1 2 +\epsilon}s^{\alpha\left(\frac1 2 +\epsilon \right)}\qquad & \mbox{if}\quad -\frac 1 2 <\epsilon<+\frac 1 2,\\
1-\frac1 2 \ln\left(\frac{1}{2bs^{\alpha}}\right)bs^{\alpha} \qquad & \mbox{if}\quad\epsilon=+\frac 1 2,\\
1-\frac{2\epsilon}{2\epsilon-1}bs^{\alpha}\qquad & \mbox{if}\quad+\frac 1 2 <\epsilon<+1,
\end{cases}
\end{equation}
and (thanks to Tauberian theorems):

\begin{equation}
f(t)\sim \begin{cases}
2^{\frac 1 2-\epsilon} \frac{\Gamma\left(\frac 1 2-\epsilon\right)\Gamma(1+\epsilon)}{\Gamma(1-\epsilon)\Gamma\left(\frac 1 2+\epsilon\right)}\frac{\alpha\left(\frac 1 2 +\epsilon\right)}{\Gamma\left(1-\frac{\alpha}{2}-\alpha\epsilon\right)}\frac{b^{\frac 1 2+\epsilon}}{t^{1+\alpha\left(\frac 1 2 +\epsilon \right)}}\qquad & \mbox{if}\quad -\frac 1 2 <\epsilon<+\frac 1 2,\\
\frac{b}{2} \ln\left(\frac{t^{\alpha}}{2b}\right)\frac{\alpha}{\Gamma(1-\alpha)}\frac{1}{t^{1+\alpha}} \qquad & \mbox{if}\quad\epsilon=+\frac 1 2,\\
\frac{2\epsilon}{2\epsilon-1}\frac{\alpha}{\Gamma(1-\alpha)}\frac{b}{t^{1+\alpha}}\qquad & \mbox{if}\quad+\frac 1 2 <\epsilon<+1.
\end{cases}
\end{equation}

Actually, we can not directly apply Tauberian theorems to $\hat{f}(s)$, of course, but we can however get around the problem by using the following trick. If we consider $\hat{f}(s)\sim 1-bs^{\eta}L\left(\frac 1 s \right)$ with $0<\eta<1$ and use an auxiliary function, for instance the derivative, then:

\begin{equation}
\hat{f}'(s)=-\int_0^{\infty}e^{-st}tf(t)dt\sim-\eta bs^{\eta-1}L\left(\frac 1 s \right)\qquad \implies \qquad tf(t)\sim \frac{\eta b}{\Gamma(1-\eta)}\frac{L(t)}{t^{\eta}}.
\end{equation}

Though we have yet to determine the result for $\epsilon \leq -\frac 1 2$. The limiting case $\epsilon=-\frac 1 2$ is almost immediate since again:

\begin{equation}
\hat{f}'(s)\sim -\frac{4\alpha}{\ln^2\left(\frac{1}{2bs^{\alpha}}\right)}\frac 1 s\quad\mbox{for}\quad s \to 0\qquad\implies\qquad f(t)\sim \frac{4\alpha}{\ln^2\left(\frac{t^{\alpha}}{2b} \right)}\frac 1 t\quad \mbox{for}\quad t \to \infty.
\end{equation}

For transient processes, instead, we must first supplement the asymptotic expansion with higher order terms:
\begin{eqnarray}
\hat{p}(s) &\sim& \frac{b}{s^{1-\alpha}}\frac{G_NF_{G_N}+(2bs^{\alpha})^{-\frac 1 2-\epsilon}K_NF_{K_N}}{G_DF_{G_D}+(2bs^{\alpha})^{+\frac 1 2-\epsilon}K_DF_{K_D}}\\
&\sim& \frac{b}{s^{1-\alpha}}\frac{G_N}{G_D}\left[ 1+(2b s^{\alpha})^{-\frac 1 2-\epsilon}\frac{K_N}{G_N}\right]\left[1-(2b s^{\alpha})^{\frac 1 2-\epsilon}\frac{K_D}{G_D}\right]\\
&\sim&\frac{b}{s^{1-\alpha}}\frac{\epsilon}{2\epsilon+1}\left[ 1+(2b s^{\alpha})^{-\frac 1 2-\epsilon}\frac{K_N}{G_N}\right],
\end{eqnarray}
in such a way that:

\begin{eqnarray}
\hat{f}(s) &\sim& 1-\frac{2\epsilon+1}{\epsilon}+\frac{2\epsilon+1}{\epsilon}\frac{K_N}{G_N}(2b s^{\alpha})^{-\frac 1 2-\epsilon},\quad \mbox{with}\quad \lim_{\epsilon\to -1^+}\hat{f}(s)=0,\: \lim_{\epsilon\to -\frac 1 2^-}\hat{f}(s)=1,\\
\hat{f}'(s)&\sim& -\frac{2\epsilon+1}{\epsilon}2^{1+2\epsilon}\frac{\Gamma\left(\epsilon+\frac1 2\right)\Gamma(-\epsilon)}{\Gamma(\epsilon+1)\Gamma\left(-\frac1 2-\epsilon\right)}\alpha\left(\frac 1 2 + \epsilon \right)(2b)^{-\frac 1 2-\epsilon}s^{-1-\alpha\left(\frac 1 2 +\epsilon\right)},\\
f(t)&\sim& 2^{\epsilon-\frac 1 2}\left( \frac{2\epsilon+1}{\epsilon}\right)^2\frac{\Gamma\left(\frac 3 2 +\epsilon\right)\Gamma(1-\epsilon)}{\Gamma(\epsilon+1)\Gamma\left(\frac 1 2-\epsilon\right)}\frac{\alpha}{\Gamma\left(1+\alpha\left(\frac 1 2 +\epsilon\right)\right)}\frac{b^{-\frac 1 2-\epsilon}}{t^{1-\frac{\alpha}{2}-\alpha\epsilon}}.
\end{eqnarray}

\subsection{First-hitting}
Now, keeping in mind the techniques illustrated in the previous Appendices~\ref{subapp-first-ret} and \ref{app-exact-ret}, we manage to generalize the results to the first-passage time to the origin, assuming to start from any other site $j_0\neq 0$.

As long as $\epsilon\neq \pm \frac 1 2$, we can write:

\begin{equation}
\hat{f}(j_0,0;\: s)=\frac{\hat{p}(j_0,0;\:s)}{\hat{p}(s)}\sim\frac{\Gamma(1+\epsilon+|j_0|)}{|j_0|!2^{|j_0|}\Gamma(1+\epsilon)}(1-|j_0|bs^{\alpha})\frac{G_NF_{G_N}+(2bs^{\alpha})^{-\frac 1 2-\epsilon}K_NF_{K_N}}{G_DF_{G_D}+(2bs^{\alpha})^{-\frac 1 2 -\epsilon}K_DF_{K_D}}.
\end{equation}

When $\epsilon \in \left(\frac 1 2,1\right)$:

\begin{equation}
\hat{f}(j_0,0;\: s)\sim\frac{\Gamma(1+\epsilon+|j_0|)}{|j_0|!2^{|j_0|}\Gamma(1+\epsilon)}(1-|j_0|bs^{\alpha})\frac{K_N}{K_D}\frac{F_{K_N}(s\to 0)}{F_{K_D}(s\to 0)}=(1-|j_0|bs^{\alpha})\frac{1+\frac{a_{K_N}b_{K_N}}{c_{K_N}}2bs^{\alpha}}{1+\frac{a_{K_D}b_{K_D}}{c_{K_D}}2bs^{\alpha}},
\end{equation}
since the first term in the expansion of hypergeometric functions ${}_2F_1(a,b;\:c;\:z)=\sum_{n=0}^{\infty}\frac{(a)_n(b)_n}{(c)_n}\frac{z^n}{n!}$ is of order $s^{\alpha}$, which is dominant with respect to $s^{\alpha\left(\frac 12+\epsilon\right)}$. Therefore:

\begin{eqnarray}
\hat{f}(j_0,0;\:s) &\sim& 1-\left[|j_0|+ \frac{(|j_0|+1-\epsilon)(|j_0|-\epsilon)}{2\epsilon-1}+\frac{\epsilon(1-\epsilon)}{2\epsilon-1}\right]bs^{\alpha}=1-\frac{j_0^2}{2\epsilon-1}bs^{\alpha},\\
f(j_0,0,t) &\sim& \frac{j_0^2}{2\epsilon-1}\frac{\alpha}{\Gamma(1-\alpha)}\frac{b}{t^{\alpha+1}}\sim\frac{j_0^2}{2\epsilon}f(t).
\end{eqnarray}

If $\epsilon\in \left(-\frac 1 2, \frac 1 2\right)$:

\begin{eqnarray}
\hat{f}(j_0,0;\:s) &\sim& (1-|j_0|bs^{\alpha})\frac{1+\frac{G_N}{K_N}(2bs^{\alpha})^{\frac 1 2+\epsilon}}{1+\frac{G_D}{K_D}(2bs^{\alpha})^{\frac 1 2+\epsilon}}\sim1-2^{\frac 1 2 +\epsilon}\left[-\frac{G_N}{K_N}+\frac{G_D}{K_D}\right]b^{\frac 1 2+\epsilon}s^{\alpha\left(\frac 1 2+\epsilon\right)},\\
f(j_0,0,t) &\sim& \frac{2^{\frac 1 2 +\epsilon}}{2^{1+2\epsilon}}\frac{\Gamma\left(-\epsilon-\frac 1 2\right)}{\Gamma\left(\epsilon+\frac 1 2\right)}\left[\frac{\Gamma(1+\epsilon)}{\Gamma(-\epsilon)}-\frac{\Gamma(1+\epsilon+|j_0|)}{\Gamma(|j_0|-\epsilon)}\right]\frac{\alpha \left(\frac1 2+\epsilon\right)b^{\frac 1 2 +\epsilon}t^{-1-\alpha\left(\frac 1 2 +\epsilon\right)}}{\Gamma\left(1-\alpha\left (\frac 1 2 +\epsilon\right)\right)}\\
&\sim& \frac{1}{2\epsilon+1}\left[\epsilon+\frac{\Gamma(1+\epsilon+|j_0|)\Gamma(1-\epsilon)}{\Gamma(1+\epsilon)\Gamma(|j_0|-\epsilon)}\right]f(t).
\end{eqnarray}

Finally, for $\epsilon\in \left(-1,-\frac 1 2\right)$:

\begin{eqnarray}
\hat{f}(j_0,0;\:s) &\sim& \frac{\Gamma(1+\epsilon+|j_0|)}{|j_0|!2^{|j_0|}\Gamma(1+\epsilon)}(1-|j_0|bs^{\alpha})\frac{G_N}{G_D}\frac{1+\frac{K_N}{G_N}(2bs^{\alpha})^{-\frac 1 2-\epsilon}}{1+\frac{K_D}{G_D}(2bs^{\alpha})^{-\frac 1 2-\epsilon}}\\
&\sim&\frac{\Gamma(1+\epsilon+|j_0|)\Gamma(-\epsilon)}{\Gamma(|j_0|-\epsilon)\Gamma(1+\epsilon)}\left[1-\left(-\frac{K_N}{G_N}+\frac{K_D}{G_D}\right)(2b)^{-\frac 1 2-\epsilon}s^{-\alpha\left(\frac 1 2+\epsilon\right)}\right],\\
f(j_0,0,t) &\sim& \frac{\alpha2^{\frac 1 2 + \epsilon}}{\Gamma\left(1+\frac{\alpha}{2}+\alpha\epsilon\right)}\frac{\Gamma\left(\frac 3 2+\epsilon\right)\Gamma(-\epsilon)}{\Gamma\left(-\epsilon-\frac 1 2\right)\Gamma(1+\epsilon)}\left[1-\frac{\Gamma(-\epsilon)\Gamma(\epsilon+1+|j_0|)}{\Gamma(|j_0|-\epsilon)\Gamma(1+\epsilon)}\right]\frac{b^{-\frac 1 2-\epsilon}}{t^{1-\alpha\epsilon-\frac{\alpha}{2}}}\\
&\sim& \frac{1}{2\epsilon+1}\left[\epsilon+\frac{\Gamma(1+\epsilon+|j_0|)\Gamma(1-\epsilon)}{\Gamma(1+\epsilon)\Gamma(|j_0|-\epsilon)}\right]f(t).
\end{eqnarray}

At this stage, we have to focus on the transition points. Firstly, let us consider $\epsilon=+\frac 1 2$:

\begin{eqnarray}
\hat{f}(j_0,0;\:s) &\sim&\frac{\Gamma\left(\frac 3 2 +|j_0|\right)}{|j_0|!2^{|j_0|}\Gamma\left(\frac 3 2\right)}(1-|j_0|bs^{\alpha})\frac{{}_2F_1\left(\frac 3 4 +\frac{|j_0|}{2}, \frac 5 4 + \frac{|j_0|}{2};\: |j_0|+1;\:\hat{\psi}^2(s)\right)}{{}_2F_1\left(\frac 5 4, \frac 3 4 ;\: 1;\:\hat{\psi}^2(s)\right)} \\
&=& \frac{\Gamma\left(\frac 3 2 +|j_0|\right)}{|j_0|!2^{|j_0|}\Gamma\left(\frac 3 2\right)}(1-|j_0|bs^{\alpha})\frac{{}_2F_1(a_N,b_N;\:a_N+b_N-1;\:\hat{\psi}^2(s))}{{}_2F_1(a_D,b_D;\: a_D+b_D-1;\:\hat{\psi}^2(s))},
\end{eqnarray}
where the hypergeometric functions in the numerator and denominator asymptotically behave as:

\begin{eqnarray}
_2F_1(a,b;\:a+b-1;\:z)&\sim& \frac{\Gamma(a+b-1)}{\Gamma(a)\Gamma(b)}\frac{1}{1-z}\left[ 1-(a-1)(b-1)\ln\left(\frac {1}{1-z}\right)(1-z)\right],\\
N &\sim&\frac{\Gamma(|j_0|+1)}{2^{\frac 1 2 -|j_0|}\sqrt{\pi}\Gamma\left(\frac 3 2 +|j_0|\right)}\frac{1}{bs^{\alpha}}\left[1-\left(\frac{j_0^2}{2}-\frac {1}{8} \right)\ln\left(\frac{1}{2bs^{\alpha}}\right)bs^{\alpha}\right],\\
D&\sim& \frac{1}{2^{-\frac 1 2}\pi b s^{\alpha}}\left[ 1+\frac 1 8 \ln\left(\frac{1}{2bs^{\alpha}}\right)bs^{\alpha}\right].
\end{eqnarray}

In conclusion:

\begin{eqnarray}
\hat{f}(j_0,0;\:s) &\sim&1-\frac{j_0^2}{2} \ln\left(\frac{1}{2bs^{\alpha}}\right)bs^{\alpha},\\
f(j_0,0,t) &\sim& \frac{j_0^2}{2}\ln\left(\frac{t^{\alpha}}{2b}\right)\frac{\alpha}{\Gamma(1-\alpha)}\frac{b}{t^{\alpha+1}}\sim j_0^2f(t).
\end{eqnarray}

Secondly, when $\epsilon=-\frac 1 2$ we get:

\begin{eqnarray}
\hat{f}(j_0,0;\:s) &\sim&\frac{\Gamma\left(\frac 1 2 +|j_0|\right)}{|j_0|!2^{|j_0|}\sqrt{\pi}}(1-|j_0|bs^{\alpha})\frac{{}_2F_1\left(\frac 1 4 +\frac{|j_0|}{2}, \frac{|j_0|}{2}+\frac 3 4;\: |j_0|+1;\:\hat{\psi}^2(s)\right)}{{}_2F_1\left(\frac 3 4, \frac 1 4 ;\: 1;\:\hat{\psi}^2(s)\right)} \\
&=&\frac{\Gamma\left(\frac 1 2 +|j_0|\right)}{|j_0|!2^{|j_0|}\sqrt{\pi}}(1-|j_0|bs^{\alpha})\frac{{}_2F_1(a_N,b_N;\:a_N+b_N;\:\hat{\psi}^2(s))}{{}_2F_1(a_D,b_D;\: a_D+b_D;\:\hat{\psi}^2(s))},
\end{eqnarray}
and as we did before:

\begin{eqnarray}
N &\sim& \frac{\Gamma(|j_0|+1)}{2^{\frac 1 2-\epsilon}\sqrt{\pi}\Gamma\left(\frac 1 2 +|j_0|\right)}\ln\left(\frac{1}{2bs^{\alpha}}\right)\left[1-\frac{2\Psi(1)-\Psi\left(\frac 1 4+\frac{|j_0| }{2}\right)-\Psi\left(\frac 3 4+\frac{|j_0| }{2}\right)}{\ln\left(\frac{1}{2bs^{\alpha}}\right)}\right],\\
D &\sim& \frac{1}{\sqrt{2}\pi}\ln\left(\frac{1}{2bs^{\alpha}}\right)\left[1-\frac{2\Psi(1)-\Psi\left(\frac 1 4\right)-\Psi\left(\frac 3 4\right)}{\ln\left(\frac{1}{2bs^{\alpha}}\right)} \right].\\
\end{eqnarray}

Hence:

\begin{eqnarray}
\hat{f}(j_0,0;\:s) &\sim& 1-\left[\Psi\left(\frac 1 4 \right) + \Psi \left(\frac 3 4 \right)-\Psi\left(\frac 1 4 +\frac{|j_0|}{2}\right)-\Psi\left(\frac 3 4 +\frac{|j_0|}{2}\right)\right]\frac{1}{\ln\left(\frac{1}{2\tau^{\alpha}s^{\alpha}}\right)},\\
f(j_0,0,t) &\sim& \left[\Psi\left(\frac 1 4 \right) + \Psi \left(\frac 3 4 \right)-\Psi\left(\frac 1 4 +\frac{|j_0|}{2}\right)-\Psi\left(\frac 3 4 +\frac{|j_0|}{2}\right)\right]\frac{\alpha}{\ln^2\left(\frac{t^{\alpha}}{2b}\right)}\frac 1 t\\
&\sim&\frac 1 4 \left[\Psi\left(\frac 1 4 \right) + \Psi \left(\frac 3 4 \right)-\Psi\left(\frac 1 4 +\frac{|j_0|}{2}\right)-\Psi\left(\frac 3 4 +\frac{|j_0|}{2}\right)\right] f(t).
\end{eqnarray}


\section{CTRW on $\mathbbm{Z}$}\label{app-CTRW-example}
Let us deal with a simple symmetric random walk, namely a nearest-neighbour (homogeneous and symmetric) random walk on the one-dimensional integer lattice, starting from the origin and ruled by waiting periods on the sites. We know \cite{First Steps in Random Walks} that the generating functions of the discrete-time model are:

\begin{equation}
P(z)=\frac{1}{\sqrt{1-z^2}},\qquad F(z)=1-\frac{1}{P(z)}=1-\sqrt{1-z^2}.
\end{equation}

Giving up the parametrization by the number of steps, thanks to Eq.~\eqref{prob-being} and Eq.~\eqref{prob-first} we can write also the Laplace transforms:

\begin{equation}
\hat{p}(s)=\frac{1-\hat{\psi}(s)}{s}\frac{1}{\sqrt{1-\hat{\psi}^2(s)}},\qquad \hat{f}(s)=1-\sqrt{1-\hat{\psi}^2(s)}.
\end{equation}

Moreover, we can recast these formulas in the required convenient way:

\begin{eqnarray}
\hat{p}(s) &=& \frac{[1-\hat{\psi}(s)]^{1-\frac 1 2}}{s}\frac{1}{\sqrt{1+\hat{\psi}(s)}}=\frac{[1-\hat{\psi}(s)]^{1-\frac 1 2}}{s}H\left(\frac{1}{1-\hat{\psi}(s)}\right),\\
\hat{f}(s)&=&1-[1-\hat{\psi}(s)]^{\frac 1 2}\sqrt{1+\hat{\psi}(s)}=1-[1-\hat{\psi}(s)]^{\frac 1 2}L\left(\frac{1}{1-\hat{\psi}(s)}\right),
\end{eqnarray}
where $L(x)=\sqrt{2-\frac{1}{x}}$ and $H(x)=\sqrt{\frac{x}{2x-1}}=1/L(x)$ are slowly-varying functions at infinity. If we further suppose, as in Section~\ref{sec-inf-mean}, that $\hat{\psi}(s)=1-b s^{\alpha}+o(s^{\alpha})$, then:

\begin{eqnarray}
p(t) &\sim& \frac{1}{\Gamma\left(1-\frac{\alpha}{2}\right)}\frac{b^{1/2}}{t^{\alpha/2}}\bar{H}(t),\\
f(t) &\sim& \frac{\alpha}{2\Gamma\left(1-\frac{\alpha}{2}\right)}\frac{b^{1/2}}{t^{1+\alpha/2}}\bar{L}(t),
\end{eqnarray}
with $\bar{L}(t)=L\left(t^{\alpha}/b\right)=\sqrt{2-bt^{-\alpha}}=1/\bar{H}(t)$.

As a final point, let us emphasize that these are acknowledged results in the literature, verified with different techniques. See, for instance, \cite{First Steps in Random Walks}.



\begin{thebibliography}{999}
\bibitem{RWRE I}
Hughes, B.D. {\em Random walks and random environments. Volume I: Random Walks.}; Clarendon Press: Oxford, UK, 1995.
\bibitem{NHRW}
Menshikov, M.; Popov, S.; Wade, A. {\em Non-homogeneous random walks. Lyapunov function methods for near-critical stochastic systems.}; Cambridge University Press: Cambridge, UK, 2017.
\bibitem{Anomalous Transport 1}
Klages, R.; Radons, G.; Sokolov I.M. {\em Anomalous Transport: Foundations and Applications.}; Wiley--VHC: Berlin, Germany, 2008.
\bibitem{Phys. Rep. 1}
Bouchaud, J.P.; Georges, A. Anomalous diffusion in disordered media: Statistical mechanisms, models and physical applications. {\em Phys. Rep.} {\bf 1990}, {\em 195}, 127–293. 
\bibitem{Nature 1}
Barthelemy, P.; Bertolotti, J.; Wiersma D.S. A Lévy flight for light. {\em Nature} {\bf 2008}, {\em453}, 495--498.
\bibitem{Adv Water Resour 1}
Dentz, M.; Cortis, A.; Scher, H.; Berkowitz, B. Time behavior of solute transport in heterogeneous media: transition from anomalous to normal transport. {\em Adv. Water Resour.} {\bf 2004}, {\em 27(2)}, 155-173.
\bibitem{Soil Sci Soc Am J 1}
Cortis, A.; Berkowitz, B. Anomalous transport in “classical” soil and sand columns. {\em Soil. Sci. Soc. Am. J. } {\bf 2004}, {\em 68(5)}, 1539-1548.
\bibitem{PLoS ONE 1}
K\"uhn, T.; Ihalainen, T.O.; Hyväluoma, J.; Dross, N.; Willman, S.F.; Langowski, J.; Vihinen-Ranta, M.; Timonen, J. Protein Diffusion in Mammalian Cell Cytoplasm. {\em PLoS ONE} {\bf 2011}, {\em 6}, e22962.
\bibitem{Phys. Rev. Lett. 2}
Nissan, A.; Berkowitz, B. Inertial Effects on Flow and Transport in Heterogeneous Porous Media. {\em Phys. Rev. Lett.} {\bf 2018}, {\em 120}, 054504.
\bibitem{Phys. Today 2}
Barkai, E.; Garini, Y.; Metzler, R. Strange kinetics of single molecules in living cells. {\em Phys. Today} {\bf 2012}, {\em 65}, 29-35.
\bibitem{Phys. Chem. Chem. Phys. 1}  
Metzler, R.; Jeon, J.H.; Cherstvy, A.G.; Barkai, E. Anomalous diffusion models and their properties: non-stationarity, non-ergodicity, and ageing at the centenary of single particle tracking. {\em Phys. Chem. Chem. Phys.} {\bf 2014}, {\em 16}, 24128-24164.
\bibitem{Phys. Rev. Lett. 3}
He, Y.; Burov, S.; Metzler, R.; Barkai, E. Random Time-Scale Invariant Diffusion and Transport Coefficients. {\em Phys. Rev. Lett.} {\bf 2008}, {\em 101}, 058101.
\bibitem{Phys. Chem. Chem. Phys. 2}
Burov, S.; Jeon, J.H.; Metzler, R.; Barkai, E. Single particle tracking in systems showing anomalous diffusion: the role of weak ergodicity breaking. {\em Phys. Chem. Chem. Phys.} {\bf 2010}, {\em 13}, 1800-1812.
\bibitem{Quart. J. Math. 1}
Gillis, J. Centrally biased discrete random walk. {\em Quart. J. Math.} {\bf 1956}, {\em 7}, 144--152.
\bibitem{ADV APPL PROBAB 1}
Percus, O.E. Phase transition in one-dimensional random walk with partially reflecting boundaries. {\em Adv. Appl. Probab.} {\bf 1985}, {\em 17(3)}, 594-606.
\bibitem{J. Math. Phys. 2}
Montroll, E.W. Random Walks on Lattices. III. Calculation of First‐Passage Times with Application to Exciton Trapping on Photosynthetic Units. {\em J. Math. Phys. } {\bf 1969}, {\em 10}, 153-165. 
\bibitem{J. Appl. Probab. 1}
Hill, J.M.; Gulati, C.M. The random walk associated by the game of roulette. {\em J. Appl. Probab.} {\bf 1981}, {\em 18}, 931-936.
\bibitem{J. Stat. Phys. 4}
Hughes, B.D.; Sahimi, M. Random walks on the Bethe lattice. {\em J. Stat. Phys.} {\bf 1982}, {\em 23}, 1688-1692.
\bibitem{Rev. Mod. Phys. 1}
Zaburdaev, V.; Denisov, S.; Klafter, J. Lévy walks. {\em Rev. Mod. Phys.} {\bf 2015}, {\em 87 (2)}, 483.
\bibitem{On growth and form 1}
Shlesinger, M.F.; Klafter J. Lévy Walks Versus Lévy Flights. In {\em On Growth and Form. Fractal and Non--Fractal Patterns in Physics.}; Stanley, H.E.; Ostrowsky, N.; Martinus Nijhoff Publihers: Dordrecht/Boston/Lancaster, USA, 1986; NATO ASI Series E: Applied Sciences - No. 100; pp. 279--283.
\bibitem{J. Math. Phys. 1}
Montroll, E.W.; Weiss, G.H. Random walks on lattices, II. {\em J. Math. Phys.} {\bf 1965}, {\em 6}, 167–181.
\bibitem{First Steps in Random Walks}
Klafter, J.; Sokolov, I.M. {\em First Steps in Random Walks. From Tools to Applications.}; Oxford University Press Inc.: New York, United States, 2011; pp. 1-51.
\bibitem{Physica A 1}
Scalas, E. The application of continuous-time random walks in finance and economics. {\em Physica A} {\bf 2006}, {\em 362}, 225--239.
\bibitem{Springer 1}
Wolfgang, P.; Baschnagel, J. {\em Stochastic Processes. From Physics to Finance}; Springer International Publishing, 2013.
\bibitem{Phys. Today 1}
Scher, H.; Shlesinger, M.F.; Bendler, J.T. Time-scale invariance in transport and relaxation. {\em Phys. Today} {\bf 1991}, 26--34.
\bibitem{Rev. Geophys. 1}
Berkowitz, B.; Cortis, A.; Dentz, M.; Scher, H. Modeling non-Fickian transport in geological formations as a continuous time random walk. {\em Rev. Geophys.} {\bf 2006}, {\em 44(2)}, RG2003.
\bibitem{Water Resour. Res. 1}
Boano, F.; Packman, A.I.; Cortis, A.; Ridolfi, R.R.L. A continuous time random walk approach to the stream transport of solutes. {\em Water Resour. Res.} {\bf 2007}, {\em 43}, W10425.
\bibitem{Water Resour. Res. 2}
Geiger, S.; Cortis, A.; Birkholzer, J.T. Upscaling solute transport in naturally fractured porous media with the continuous time random walk method. {\em Water Resour. Res.} {\bf 2010}, {\em 46(12)}, W12530.
\bibitem{J. Stat. Mech. 1}
Onofri, M.; Pozzoli, G.; Radice, M.; Artuso, R. Exploring the Gillis model: a discrete approach to diffusion in logarithmic potentials. {\em J. Stat. Mech.} {\bf 2020}, {\em 2020}, 113201.
\bibitem{Cambridge 1}
Redner, S. {\em A Guide to First-Passage Processes}; Cambridge University Press: Cambridge, UK, 2001. 
\bibitem{Handbook 1}
Abramowitz, M.; Stegun, I.A. {\em Handbook of mathematical functions}; Dover Publications Inc: New York, USA, 1970.
\bibitem{Physica A 2}
Hughes, B.D. On returns to the starting site in lattice random walks. {\em Physica A} {\bf 1986}, {\em 134}, 443-457.
\bibitem{Physica D 1}
Castiglione, P.; Mazzino, A.; Muratore-Ginanneschi, P.; Vulpiani, A. On strong anomalous diffuusion. {\em Physica D} {\bf 1999}, {\em 134}, 75-93.
\bibitem{arXiv 1}
Janson, S. Stable Distributions. {\em arXiv:1112.0220v2 [math.PR]} {\bf2011} (lecture notes).
\bibitem{Feller II}
Feller, W. {\em An Introduction to Probability Theory and Its Applications. Vol. II}; John Wiley and Sons, Inc.: New York, USA, 1971.
\bibitem{Math. Scand. 1}
Sparre Andersen, E. On the fluctuations of sums of random variables II. {\em Math. Scand.} {\bf 1954}, {\em 2}, 195-223.
\bibitem{J. Phys. A Math. Theor. 1}
Mounaix, P.; Majumdar, S.N.; Schehr, G. Statistics of the number of records for random walks and Lévy Flights on a 1D Lattice. {\em J. Phys. A Math. Theor.} {\bf 2020}, {\em 53}, 415003.
\bibitem{Feller I}
Feller, W. {\em An Introduction to Probability Theory and Its Applications. Vol. I}; John Wiley and Sons, Inc.: New York, USA, 1968.
\bibitem{Phys. Rev. E 2}
Artuso, R.; Cristadoro, G.; Degli Esposti, M.; Knight, G.S. Sparre-Andersen theorem with spatiotemporal correlations. {\em Phys. Rev. E} {\bf 2014}, {\em 89}, 052111.
\bibitem{Phys. Rev. E 1}
Radice, M; Onofri, M.; Artuso, R.; Pozzoli, G. Statistics of occupation times and connection to local properties of nonhomogeneous random walks, {\em Phys. Rev. E} {\bf 2020}, {\em 101}, 042103.
\bibitem{Trans. Am. Math. Soc. 1}
Darling, D.A.; Kac, M. On occupation times for Markoff processes. {\em Trans. Am. Math. Soc.} {\bf 1957}, {\em 84}, 444-458.
\bibitem{Trans. Am. Math. Soc. 2}
Lamperti, J. An occupation time theorem for a class of stochastic processes. {\em Trans. Am. Math. Soc.} {\bf 1958}, {\em 88}, 380-387.
\bibitem{J. Stat. Phys. 2}
Godrèche, C.; Luck, J.M. Statistics of the occupation time for renewal processes. {\em J. Stat. Phys.} {\bf 2001}, {\em 104}, 489-524.
\bibitem{J. Stat. Phys. 3}
Barkai, E. Residence time statistics for normal and fractional diffusion in a force field. {\em J. Stat. Phys.} {\bf 2006}, {\em 123}, 883-907.
\bibitem{J. Phys. Condens. Matter 1}
Bel, G.; Barkai, E. Occupation time and ergodicity breaking in biased continuous time random walks. {\em J. Phys. Condens. Matter} {\bf 2005}, {\em 17}, S4287-S4304.
\bibitem{Phys. Rev. E 3}
Bel, G.; Barkai, E. Random walk to a nonergodic equilibrium concept. {\em Phys. Rev. E} {\bf 2006}, {\em 73}, 016125.
\bibitem{Phys. Rev. Lett. 1}
Bel, G.; Barkai, E. Weak ergodicity breaking in continuous-time random walk. {\em Phys. Rev. Lett.} {\bf 2005}, {\em 94}, 240602.
\bibitem{Princeton 1}
Widder, D.V. {\em The Laplace transform}; Princeton University Press: London, UK, 1946.
\bibitem{J. Stat. Phys. 1}
Dechant, A.; Lutz, E.; Barkai, E.; Kessler, D.A. Solution of the Fokker-Planck Equation with a Logarithmic Potential. {\em J. Stat. Phys.} {\bf 2011}, {\em 145}, 1524-1545.
\bibitem{Stoch. Process. Their Appl. 1}
Hryniv, O.; Menshikov, M.V.; Wade, A.R. Excursions and path functionals for stochastic processes with asymptotically zero drifts. {\em Stoch. Process. Their Appl.} {\bf 2013}, {\em 123}, 1891-1921.
\end{thebibliography}
\end{document}